\documentclass[referee,useAMS,usenatbib]{mn2e}
\usepackage{graphicx,graphics,amsmath,amssymb,mathrsfs}
\usepackage{subeqn, wasysym}

\newcommand{\beq}{\begin{equation}}
\newcommand{\eeq}{\end{equation}}

\newcommand{\bfr}{\mbox{\boldmath $r$}}
\newcommand{\bfA}{\mbox{\boldmath $A$}}
\newcommand{\bfX}{\mbox{\boldmath $X$}}
\newcommand{\bfu}{\mbox{\boldmath $u$}}

\newcommand{\cross}{\mbox{\boldmath $\times$}}
\newcommand{\cendot}{\mbox{\boldmath $\cdot$}}
\newcommand{\p}{\mbox{$\partial$}}
\newcommand{\tk}{\mbox{$T_{\rm kep}$}}
\newcommand{\ts}{\mbox{$T_{\rm sec}$}}

\newcommand{\scrd}{{\cal D}}
\newcommand{\scri}{{\cal I}}
\newcommand{\scrr}{{\cal R}}

\begin{document}

\title[Stellar Dynamics around a Massive Black Hole~I]{Stellar Dynamics around a Massive Black Hole~I:\\Secular Collisionless Theory}

\author[Sridhar \& Touma]{S.~Sridhar$^{1,3}$ and Jihad~R.~Touma$^{2,4}$\\
$^{1}$~Raman Research Institute, Sadashivanagar, Bangalore 560 080, India\\
$^{2}$~Department of Physics, American University of Beirut, PO Box 11-0236, Riad El-Solh, Beirut 11097 2020, Lebanon\\  
$^{3}$~ssridhar@rri.res.in $\quad^{4}$~jt00@aub.edu.lb\\
}
\maketitle
\vspace{-2em}
\begin{abstract}
We present a theory in three parts, of the secular dynamics of a (Keplerian) stellar system of mass $M$ orbiting a black hole of mass $M_\bullet \gg M$. Here we describe the collisionless dynamics; Papers~II and III are on the (collisional) theory of Resonant Relaxation. The mass ratio, $\varepsilon = M/M_\bullet \ll 1$, is a natural small parameter implying a separation of time scales between the short Kepler orbital periods and the longer orbital precessional periods. The collisionless Boltzmann equation (CBE) for 
the stellar distribution function (DF) is averaged over the fast Kepler orbital phase using the method of multiple scales. The orbit--averaged system is described by a secular DF, $F$, in a reduced phase space. $F$ obeys a secular CBE that includes stellar self--gravity, general relativistic corrections up to 1.5 post--Newtonian order, and external sources varying over secular times. Secular dynamics, even with general time dependence, conserves the semi--major axis of every star. This additional integral of motion promotes extra regularity of the stellar orbits, and enables the construction of equilibria, $F_0$, through a secular Jeans theorem. A linearized secular CBE determines the response and stability of $F_0$. Spherical, non--rotating equilibria may support long--lived, warp--like distortions. We also prove that an axisymmetric, zero--thickness, flat disc is secularly stable to all in--plane perturbations, when its DF, $F_0$, is a monotonic function of the angular momentum at fixed energy.
\end{abstract}


\section{Introduction}

Massive black holes (MBHs), consuming gas and stars, are believed to be responsible for the radiant power of quasars and other active galactic nuclei \citep{ree84,kro99}. There is also dynamical evidence for MBHs with masses $M_\bullet$ in the range $10^5 - 10^{10}\,M_{\odot}\,$ in the nuclei of more nearby galaxies \citep{kh13}.  Coevolution (or otherwise) of a MBH and its host galaxy has been discussed using many studies of correlations of $M_\bullet$ with galaxy properties \citep{kh13,hb14}. 
Ground--based observations provided the first dynamical evidence for MBHs in nearby galaxies \citep{kr95}. They also revealed nuclear star clusters  where stellar number densities far exceeded those of globular clusters. Early work on the stellar dynamics of galactic nuclei considered several aspects of the interactions between the MBH and its stellar environment: feeding of stars to a MBH  through two--body relaxation \citep{bw76,ck78}; the effect of the adiabatic growth of a MBH on the formation of stellar density cusps \citep{you80}, and on stellar orbits \citep{gb84}; destruction by a MBH of the regularity of stellar orbits in the inner parts of triaxial galaxy \citep{gb85}. With the \emph{Hubble Space Telescope} spatially resolved kinematics, of the dense nuclear clusters of stars populating the MBH's sphere of influence, became available.  Recent ground--based observations with adaptive optics have resolved the sphere of influence of the MBH at our Galactic Centre so well, that the orbits of many individual stars are now being followed \citep{geg10}. 

Our interest is in these dense stellar systems in galactic nuclei lying within the sphere of influence of the MBH. Following \citet{tre05} we may consider a stellar system orbiting a MBH to be \emph{Keplerian}\footnote{\citet{tre05} uses the term ``near--Keplerian''; we have shortened it to Keplerian.} when the gravitational force, over a large range of radii, is dominated by the MBH, and the dominant non--Keplerian force arises from the self--gravity of all the stars. In the absence of self--gravity the orbit of a star follows a closed Kepler ellipse around the MBH, because the radial and azimuthal frequencies are equal to each other (i.e. they are ``degenerate'' or ``resonant''). When the small self--gravity of the cluster is taken into account, the orbit of a star is usefully represented by its instantaneous osculating Kepler ellipse, which evolves slowly over \emph{secular} time scales that are much longer than the fast orbital times around the MBH. Progress in the understanding of the dynamics of Keplerian stellar systems can perhaps be traced to two papers: \citet{tre95} on the structure of the (then) double nucleus of the Andromeda galaxy (M31), and \citet{rt96} on the role of resonant relaxation in stellar systems with degenerate (or resonant) frequencies. The former  has stimulated much work on  (collisionless) \emph{dynamics} whereas the latter proposed a new and efficient (collisional) relaxation process called Resonant Relaxation, which is the \emph{statistical mechanics} of Keplerian stellar systems.
Two decades of research has paved the way for the construction of triaxial, 
even lopsided, equilibria around MBHs, and simplified the modeling, stability analysis, and numerical simulation of nuclear star clusters. We 
now have a better appreciation of the dynamics and the statistical mechanics of Keplerian star clusters, and of the additional role of the MBH as a
consumer of stars. But the whole subject lacks a clearly formulated and general mathematical foundation. Our aim is to fill this gap and provide a \emph{secular theory of the structure and evolution of Keplerian stellar systems}, described by the slow time evolution of a secular \emph{distribution function} (DF), defined in a reduced phase space.
Here (Paper~I) we derive the (collisionless) theory describing their dynamics, beginning with the standard description of collisionless stellar systems given in \citet{bt08}. In \citet{st15} we deal with their (collisional) statistical mechanics and present a first--principles theory of resonant relaxation based on \citet{gil68}. This is then applied
in \citet{st16} to the physical kinetics of axisymmetric, zero--thickness, flat discs.

The sketch of a model offered in \citet{tre95}  --- for a lopsided Keplerian disc of stars around a MBH in the nucleus of M31 --- has since been fleshed out in a number of numerical models \citep{sta99,bac01,ss01,ss02,pt03,ss04,kt13,bm13}. This has brought to the forefront deep and interesting questions regarding the general nature of secular collisionless evolution. Below we give a brief time--line of developments in the secular \emph{collisionless dynamics} of Keplerian stellar systems. Work in this field may be broadly divided into explorations of orbital dynamics, construction of stationary DFs, linear stability of stationary DFs, and the nonlinear evolution of the DF.

\medskip 
\noindent
{\bf \emph{Orbital Dynamics}:} 
The orbital structure of model galactic--type potentials with a central MBH was studied in \citet{st97a,st97b,mv99}.
\citet{st99} imported the Gaussian orbit--averaging methods of celestial mechanics into stellar dynamics within the sphere of influence of a MBH. In this procedure a stellar orbit is averaged over the (fast--varying) orbital phase of the corresponding osculating Kepler ellipse, a direct consequence of which is the near conservation of the semi--major axis of the stellar orbit. Since there arises an extra (approximate) integral of motion, they concluded that \emph{a MBH promotes extra order on stellar orbits within their sphere of influence}. Orbit--averaging takes us to a reduced phase space describing secular dynamics: this is the theatre of the coupled evolution of the orbital angular momentum and Lenz vectors --- giving the eccentricity and spatial orientation of the orbit --- keeping fixed the semi--major axis. Studying orbits in planar, lopsided model disc potentials, they found a family of stable orbits that were lopsided in the same sense as the disc potential. The more complex orbital structure of triaxial systems was also studied in a number of papers: see e.g. \citet{ss00,pm01,mv11}. Of particular interest is a family of centrophilic orbits that bring stars close to the MBH. These are reviewed in \citet{mer13}, which also offers a treatment of the post--Newtonian (relativistic) effects of the MBH on stellar orbital dynamics. 

\medskip 
\noindent

{\bf \emph{Linear Stability of Stationary DFs}:}
Stationary DFs are often taken as first approximations for describing the nuclear star clusters, and their stability to small perturbations is an important question. This is, in general, a difficult problem because the 
long--range nature of gravitational interactions leads to an integral 
eigenvalue problem. Two types of stellar systems have been studied: (a) 
Zero--thickness flat discs, and (b) Spherical non--rotating systems. 

{\bf (a) \emph{Linear Stability of Discs}:} \citet{sst99} used the 
Laplace--Lagrange secular theory to study slow, lopsided $m=1$ modes (where $m$ is the azimuthal wavenumber) of cold stellar discs around MBHs.
\citet{lg99} considered nonlinear $m=1$ spiral density waves in fluid discs, and used a variational method to derive the dispersion relation and angular momentum fluxes. The numerical simulations of \citet{js01} provided evidence for long--lived lopsided modes. \citet{tre01} demonstrated that a variety 
(fluid, stellar and softened--gravity) of rotating discs were neutrally stable to $m=1$ modes in the Wentzel--Kramers--Brillouin (WKB) limit. 
He formulated a linear integral eigenvalue problem for softened--gravity discs, and demonstrated that all $m=1$ modes were neutrally stable; 
the problem was also solved numerically to determine the eigenfrequencies and eigenfunctions. \citet{jt12} derived the WKB dispersion relationship for slow modes in a collisionless disc with a Schwarzschild DF and showed that modes of all $m$ were stable. \citet{tou02} presented a Laplace--Lagrange theory for discs composed of softened Gaussian wires, confirmed the stability result of \citet{tre01} for prograde discs, and also demonstrated that \emph{counter--rotating discs were unstable to slow, lopsided modes}; a possible origin of the lopsided nuclear disc of M31 may be through such an instability. \citet{ss02} built on this to include a retrograde population of stars in their planar model of the nuclear disc of M31, and suggested that this population
could arise naturally through an accretion event. \citet{tre05} considered stellar discs with DFs even in the angular momentum, and empty loss cones (i.e. DF is zero at zero angular momentum). Using the variational principle of \citet{goo88} he argued that many of them are likely to be unstable to $m=1$ modes. \citet{pps07} considered mono--energetic discs dominated by nearly radial orbits, and found that they are prone to loss cone instabilities of all $m$, if there is some amount of counter--rotating stars. \citet{ss10} derived the instability of $m=1$ modes of softened--gravity, counter--rotating discs in the WKB limit. \citet{gss12} studied the linear integral eigenvalue problem for these discs and determined eigenvalues, precession frequencies and growth rates.  

{\bf (b) \emph{Linear Stability of Spherical Non--rotating Systems}:} 
\citet{tre05} formulated the eigenvalue problem for the linear secular modes of spherical non--rotating systems and determined the following: 
(i) Modes with arbitrary $l$ (where $l$ is the latitudinal wavenumber) 
are of two types --- either pairs of modes that are damped and growing in time, or pairs of oscillatory modes; (ii) Lopsided $l=1$ modes are stable 
when the DF is either a decreasing function of angular momentum (at fixed energy), or the DF is an increasing function of angular momentum and has a non--empty loss cone; (iii) DFs that are increasing functions of 
angular momentum with an empty loss cone are neutrally stable to an 
$l=1$ mode that corresponds to a uniform displacement. \citet{pps07} considered a loss--cone instability for mono--energetic spherical stellar systems dominated by nearly radial orbits. They found that DFs which are non monotonic functions of the angular momentum may be unstable to modes with $l\geq 3$ linear perturbations; a necessary condition for instability is the retrograde precession of orbits, which obtains naturally in such clusters.

\medskip
\noindent
{\bf \emph{Nonlinear evolution of the DF}:} Numerical studies of the collisionless dynamics of Keplerian stellar systems have been more limited than their galactic counterparts. This is partly because observations are not yet precise enough to finely resolve the sphere of influence of central black holes (the Galactic center and M31's nucleus being the exceptions), and partly because the kind of N--body simulations required to resolve the collisionless regime are computationally costly when a massive central body is present. Schwarzschild--type methods have been used to construct non--axisymmetric equilibria, both kinematic \citep{pt03} and self-consistent  \citep{pm01, ss02, bm13}).  Early numerical simulations \citep{bac01, js01} showed collisionless relaxation of perturbed stellar discs. \citet{ttk09} built a numerical algorithm which is particularly suited to the numerical exploration of secular clusters. This is based on a generalization of an orbit--averaging algorithm, first introduced by Gauss, to systems with Plummer--softened interactions. The simulations proceed over secular time steps, in a vectorial framework which avoids singularities in the usual Keplerian elements. Averaging of force evaluations, which is costly, is ideal for parallel computations on a Beowulf cluster with or without GPU option. Their work explored various violent regimes of secular interactions, and provides the first evidence for collisionless relaxation of secularly unstable counter--rotating clusters into lopsided configurations. The (fully nonlinear) collisionless Boltzmann equation (CBE) for secular evolution was presented in \citet{ts12}, and used to explore 
the nonlinear, collisionless dynamics of counter--rotating discs. 
\citet{kt13} performed N--body simulations of unstable counter--rotating 3--dim clusters, and demonstrated that they relax collisionlessly into triaxial lopsided configurations; these appear to be promising models of the lopsided stellar nucleus of M31.

We are now in a position to bring all the work together under a general 
framework, describing the secular collisionless evolution of Keplerian 
stellar systems. We begin in \S~2 by casting the standard collisionless Boltzmann equation for the phase space DF in the Delaunay action--angle variables (which are natural phase space coordinates for the pure Kepler problem). In \S~3 the method of multiple scales is used to derive an explicit functional form for the DF. Complementing this is Appendix~A
on canonical perturbation theory applied to an individual stellar 
orbit. In \S~4 self--gravity is augmented by relativistic corrections and external gravitational perturbations, to obtain the secular CBE governing the evolution of the secular DF. A secular Jeans theorem is proved in \S~5, and used to discuss secular equilibria. Linear secular perturbations and stability are discussed in \S~6, where some results are presented for two types of systems: (a) Spherical non--rotating systems; (b) Zero--thickness flat axisymmetric discs. Concluding remarks are offered in \S~7.

\section{Collisionless evolution of Keplerian Stellar Systems}

In a Keplerian stellar system the gravitational force over a large range of radii is dominated by the MBH, and the dominant non--Keplerian force arises from the self--gravity of all the stars. The MBH also plays an additional role in destroying stars on orbits with near--zero angular momentum, through tidal disruption or being swallowed whole. Let us consider such a system composed of a large number ($N\gg 1$) stars, of total mass $M$ and size $R$, orbiting a MBH of mass $M_\bullet$. Then we must have (1) $\varepsilon \equiv (M/M_\bullet) \ll 1\,$, and (2) $\,R \gg r_{\bullet} \equiv 2GM_\bullet/c^2$ such that $(r_{\bullet}/R) \ll \varepsilon$; this ensures that, over most of the cluster, stellar self--gravity dominates relativistic corrections to the MBH's gravity. Then $\varepsilon$ is the natural small parameter for the dynamics of non--relativistic star clusters.

The important time scales in the life of a Keplerian stellar system are: (i) The short Kepler orbital timescale, $\tk = 2\pi\left(R^3/GM_\bullet\right)^{1/2}\,$; (ii) The longer secular timescale, $\ts\,$, over which the stars behave as if they were Keplerian Ellipses being torqued collisionlessly by the mean gravitational field of the cluster; (iii) The resonant relaxation timescale, $T_{\rm res}\,$, over which the angular momenta (but not the energies) of stellar orbits diffuse due to gravitational collisions between stars; (iv) The 2--body relaxation timescale, $T_{\rm relax}$, over which both the energies and angular momenta of the stellar orbits diffuse, due to gravitational collisions between stars. The different time scales are related to each other by, $\,\tk:\ts:T_{\rm res}:T_{\rm relax} \,=\, 1: \varepsilon^{-1}: N\varepsilon^{-1}: N\varepsilon^{-2}\,$.\footnote{This estimate of $T_{\rm res}$ applies to the diffusion of the magnitude of angular momentum, or \emph{scalar} resonant relaxation. Logarithmic
corrections have been neglected in the estimate for $T_{\rm relax}$.} For galactic nuclei of interest $T_{\rm relax}$ is usually greater than the Hubble time, so the usual 2--body relaxation process is not important. However, $T_{\rm res}$ can be less than the Hubble time, so there are three regimes of interest:
\begin{itemize}
\item \emph{Short times} $\left(t \,\lesssim\, \mbox{few}\;\tk\,\ll \ts\,\right)$ during which there are 
variations in the stellar distribution over Kepler orbital times. These 
have indeed been directly observed in our Galactic Centre. The general theory is undeveloped and could be important for understanding the first 
phase of the collisionless relaxation of a star cluster, were it to form 
in the sphere of influence of the MBH. This regime is not studied in this paper.
\item \emph{Intermediate times} $\left(\tk \,\ll\, \,t\,\sim\, \mbox{few}\;\ts\, \,\ll\, T_{\rm res}\right)$ over which the cluster behaves collisionlessly. \emph{This secular regime is the focus of the present work} (Paper~I).
\item \emph{Long times} $\left(\ts \,\ll\, \,t\,\sim\, \mbox{few}\;T_{\rm res}\,\ll\, T_{\rm relax}\right)$ over which the system evolves collisionally due to resonant relaxation. This is explored in Papers~II and III.
\end{itemize}
The two Keplerian systems that have been studied best are also the 
closest ones, in which the sphere of influence of the MBH is well--resolved
\citep{kh13}. The Galactic Centre MBH has an estimated mass of $\sim 4.6 \times 10^6~M_{\odot}$ \citep{yel14}, and is surrounded by many young and old stars, with $\varepsilon \sim  10^{-2} - 10^{-1}$ depending on what substructure is being studied. The `triple nucleus' of M31 has a MBH of mass $\sim 1.4 \times 10^8~M_{\odot}$ \citep{ben05}, with a similar values of $\varepsilon$, depending on where the disc edge is located. One may ask if these values of $\varepsilon$ are small enough. For M31 the answer seems to be in the affirmative because, as discussed earlier, the Keplerian disc model provides a natural explanation of the lopsidedness in terms of the 
trapping of orbits by self--gravity. The orbits of many individual stars close to the Galactic Centre MBH have been determined, and for these the dynamics is certainly Keplerian. Disc--like structures have also been 
observed, e.g. \citet{yel14}, but secular theory has not progressed to the stage of making clear predictions that can be tested.

\subsection{Collisionless Boltzmann equation in the Delaunay variables}

Let $\bfr$ be the position of a test star relative to the MBH, 
$t$ be the time variable, and $\bfu = ({\rm d}\bfr/{\rm d}t)$ be
the relative velocity. A large $N$ stellar system can be described at 
time $t$ by $\,f(\bfr, \bfu, t)$, the \emph{distribution function} (DF)  in the 6--dim single--particle phase space $\{\bfr, \bfu\}$. The DF is positive and, when the MBH is not considered a sink of stars, the total mass of the 
stellar system is conserved. Then the DF can be normalized for all time:
\beq
\int f(\bfr, \bfu, t)\,{\rm d}\bfr\,{\rm d}\bfu \;=\; 1, 
\label{norm}
\eeq

\noindent
making $f$ a (single--particle) probability distribution function. The 
gravitational force experienced by a star is due to the MBH and all the other stars. Per unit mass the former is $\,-GM_\bullet\hat{\bfr}/r^2\,$, and the latter can be written as $\,-\p(\varepsilon\varphi)/\p \bfr\,$ where $\,\varepsilon\varphi(\bfr, t)\,$ is the mean--field self--gravitational potential given by, 
\beq
\varphi(\bfr, t) \;=\; -\,GM_\bullet\int 
\frac{f(\bfr', \bfu', t)}{\vert\bfr - \bfr'\vert}
\,{\rm d}\bfr'\,{\rm d}\bfu'\,.
\label{phiself}
\eeq
\noindent
We write the acceleration of the MBH due to all the stars as $\varepsilon\bfA_\bullet(t)$, where
\beq
\bfA_\bullet(t) \;=\; GM_\bullet\int 
f(\bfr, \bfu, t)\,\frac{\hat{\bfr}\;}{r^2}
\,{\rm d}\bfr\,{\rm d}\bfu\,.
\label{aself}
\eeq
\noindent
Note that $\varphi$ and $\bfA_\bullet$ have been defined such that they are $O(1)$ quantities. Then the relative acceleration of a star (with respect to the MBH) is obtained by subtracting the acceleration of the MBH from the 
net gravitational force (per unit mass) on the star:
\beq
\frac{{\rm d}\bfu}{{\rm d}t} \;=\; -\,\frac{GM_\bullet}{r^2}\hat{\bfr} \;-\; \varepsilon\frac{\p \varphi}{\p \bfr} \;-\; \varepsilon\bfA_\bullet\,.
\label{eomstar}
\eeq

The stellar system may be regarded as collisionless over times
shorter than the relaxation times for exchanging energy or angular momentum through stellar gravitational encounters. Then the time evolution of the DF is governed by the \emph{collisionless Boltzmann equation} (CBE):   
\begin{eqnarray}
\frac{{\rm d}f}{{\rm d}t} &\;\equiv\;& 
\frac{\p f}{\p t} \;+\; \frac{{\rm d}\bfr}{{\rm d} t}\cendot\frac{\p f}{\p \bfr} \;+\; \frac{{\rm d}\bfu}{{\rm d} t}\cendot\frac{\p f}{\p \bfu}  
\nonumber\\[1em]
&\;=\;&
\frac{\p f}{\p t} \;+\; \bfu\cendot\frac{\p f}{\p \bfr} 
\;+\; \left(-\,\frac{GM_\bullet}{r^2}\hat{\bfr} \;-\; \varepsilon\frac{\p \varphi}{\p \bfr} \;-\; \varepsilon\bfA_\bullet\right)\cendot\frac{\p f}{\p \bfu} \;=\; 0\,.
\nonumber
\end{eqnarray}

\noindent
It can be verified that the CBE preserves the normalization of eqn.(\ref{norm}). Using Poisson Brackets (PBs), the CBE can be rewritten compactly as:
\beq
\frac{\p f}{\p t} \;+\; \left[\,f\,,\,H_{\rm org}\,\right]_{(6)} \;=\; 0\,,
\label{cbe}
\eeq

\noindent
where 
\beq
H_{\rm org}(\bfr, \bfu, t) \;=\; \frac{u^2}{2} \,-\, \frac{GM_\bullet}{r} \;+\;
\varepsilon\varphi(\bfr, t) \;+\; 
\varepsilon\bfr\cendot\bfA_\bullet(t)
\label{ham0}
\eeq

\noindent 
is the Hamiltonian, and $[\;,\;]_{(6)}$ is the 6--dim PB defined by:\footnote{We have used the subscript `6' for the full PB because, 
in much of this paper, we will use a reduced PB (without subscript) --- see eqns.(\ref{pb624}) and (\ref{pbdel4}) --- that operates in a 4--dim subspace.}
\beq
\left[\,f\,,\,H_{\rm org}\,\right]_{(6)} \;=\; \frac{\p f}{\p \bfr}\cendot\frac{\p H_{\rm org}}{\p \bfu} \;-\; \frac{\p f}{\p \bfu}\cendot\frac{\p H_{\rm org}}{\p \bfr}\,.
\label{pb}
\eeq

\noindent
The Hamiltonian $H_{\rm org}$ of eqn.(\ref{ham0}) is the sum of an $O(1)$ term $\,E_{\rm k} = \left(u^2/2 - GM_\bullet/r\right)\,$ which is the 
Keplerian orbital energy, and an $O(\varepsilon)$ term $\,\varepsilon(\varphi + \bfr\cendot\bfA_\bullet)\,$ which comes from the gravity of all the stars. Then the CBE can be written as:
\beq
\frac{\p f}{\p t} \;+\; \left[\,f\,,\,E_{\rm k}\,\right]_{(6)} \;+\;
\varepsilon\left[\,f\,,\,\varphi + \bfr\cendot\bfA_\bullet\,\right]_{(6)}
\;=\; 0\,,
\label{cbe-new}
\eeq

\noindent
Since the Keplerian part is dominant, it is convenient to change the 
canonical variables from $\{\bfr, \bfu\}$ to new ones, the \emph{Delaunay variables} in which the Kepler motion of every star appears simplest. 

The Delaunay variables, $\scrd\equiv\{I, L, L_z; w, g, h\}$, are a set of action--angle variables for the Kepler problem \citep{plu60,md99,bt08}. The three actions are: $I\,=\,\sqrt{GM_\bullet a\,}\,$; $L\,=\, I\sqrt{1-e^2\,}$ the magnitude of the angular momentum; and $L_z\,=\, L\cos{i}\,$ the $z$--component of the angular momentum. The three angles conjugate to them are, respectively: $w$ the orbital phase; $g$ the angle to the periapse from the ascending node; and $h$ the longitude of the ascending node. When expressed 
in terms of the $\scrd$ variables, the Keplerian orbital energy assumes the simple and degenerate form $E_{\rm k}(I) = -1/2(GM_\bullet/I)^2$. Hence for the Kepler problem all the Delaunay variables, excepting the orbital phase $w$, are constant in time; $w$ itself advances at the rate 
\beq
\Omega_{\rm k}(I) \;=\; \frac{{\rm d} E_{\rm k}}{{\rm d} I} 
\;=\; \frac{(GM_\bullet)^2}{I^3}\,.
\label{kepfreq}
\eeq 

\noindent
In terms of the $\scrd$ variables, the 6--dim PB between 
two functions, $\chi_1(\scrd)$ and $\chi_2(\scrd)\,$, is:
\beq
\left[\,\chi_1\,,\,\chi_2\,\right]_{(6)} \;=\;
\left(\frac{\p \chi_1}{\p w}\frac{\p \chi_2}{\p I} -
\frac{\p \chi_1}{\p I}\frac{\p \chi_2}{\p w}\right) \,+\, 
\left(\frac{\p \chi_1}{\p g}\frac{\p \chi_2}{\p L} -
\frac{\p \chi_1}{\p L}\frac{\p \chi_2}{\p g}\right) \,+\, 
\left(\frac{\p \chi_1}{\p h}\frac{\p \chi_2}{\p L_z} -
\frac{\p \chi_1}{\p L_z}\frac{\p \chi_2}{\p h}\right)\,.
\label{pbdel}
\eeq

\noindent
The first PB in eqn.(\ref{cbe-new}) is $\,\left[\,f\,,\,E_{\rm k}\,\right]_{(6)} \,=\, \Omega_{\rm k}\left(\p f/\p w\right)\,$. In the second PB, $\,\varphi$ and $\,\bfA_\bullet$ have to be first expressed in terms of the $\scrd$ variables. For this we need to rewrite the phase--space integrals of eqns.(\ref{phiself}) and (\ref{aself}) in the $\scrd$ variables. The DF is $\,f(\scrd, t)$ and $\,{\rm d}\bfr\,{\rm d}\bfu = {\rm d}\scrd \equiv {\rm d}I\,{\rm d}L\,{\rm d}L_z\,{\rm d}w\,{\rm d}g\,{\rm d}h\,$. Lastly, the position vector 
$\bfr(\scrd) = (x, y, z)$ is given by \citep{plu60,md99,ss00}:
\beq
{\left( \begin{array}{c} x \\ \\ y \\ \\ z \end{array} \right) } 
 = 
{\left( \begin{array}{ccc} 
   C_{g} C_{h} - C_{i} S_{h} S_{g} & \quad -S_{g} C_{h} - C_{i} S_{h} C_{g} & 
   \quad S_{i} S_{h}  \\ \\
   C_{g} S_{h} + C_{i} C_{h} S_{g} & \quad -S_{g} S_{h} + C_{i} C_{h} C_{g} & 
   \quad -S_{i} C_{h} \\ \\
   S_{i} S_{g}               & \quad S_{i} C_{g}            & \quad C_{i}
	\end{array} \right)} {\left( \begin{array}{c} 
   a(C_{\eta} - e) \\  \\ a\sqrt{1 - e^2\,}\, S_{\eta} \\  \\ 0 \end{array} 
   \right)}
\label{xyzdel}
\eeq
\noindent 
where $S$ and $C$ are shorthand for sine and cosine of the angles given as 
subscript. Here $a$ is the semi--major axis; $\eta$ is the eccentric anomaly, related to the orbital phase through $w= (\eta - e\sin\eta)\,$; 
$\,e=\sqrt{1-L^2/I^2\,}$ is the eccentricity; and $i$ is the inclination angle determined by $\cos i = (L_z/L)\,$. It is also useful to note that
$r = \sqrt{x^2 + y^2 + z^2} = a(1 - e\cos\eta)\,$. Then eqns.(\ref{phiself}) and (\ref{aself}) take the form:
\beq
\varphi(\scrd, t) \;=\; -\,GM_\bullet\int 
\frac{\,f(\scrd', t)\,}{\,\vert\bfr - \bfr'\vert\,}
\,{\rm d}\scrd'\,,\qquad\qquad
\bfA_\bullet(t) \;=\; GM_\bullet\int 
f(\scrd, t)\,\frac{\hat{\bfr}\;}{r^2}
\,{\rm d}\scrd\,,
\label{phiadel}
\eeq
   
\noindent
where $\bfr=\bfr(\scrd)$ and $\bfr' = \bfr(\scrd')$ are given in eqn.(\ref{xyzdel}). Then the CBE of eqn.(\ref{cbe-new}) in the $\scrd$ variables is:
\beq
\frac{\p f}{\p t} \;+\;  \Omega_{\rm k}\frac{\p f}{\p w} \;+\;
\varepsilon\left[\,f\,,\,\varphi \,+\, \bfr\cendot\bfA_\bullet\,\right]_{(6)}
\;=\; 0\,.
\label{cbe3}
\eeq
This form of the CBE makes apparent two natural time scales in the problem:
(i) the Kepler orbital period $\tk = 2\pi/\Omega_{\rm k}\,$, on which the phase $w$ varies, and (ii) the longer \emph{secular} time scale $\ts = \varepsilon^{-1}\tk\,$. In the next section we \emph{orbit--average the CBE over the fast orbital phase $w$}, and obtain a reduced CBE describing secular evolution.

\section{Orbit--averaging via the Method of Multiple Scales}

Since there are two well--separated time scales $\,\left(\tk\;\, \mbox{and} \,\;\ts=\varepsilon^{-1}\tk \gg \tk\right)\,$ in the problem, it is convenient to analyze the CBE of eqn.(\ref{cbe3}) using the \emph{method of multiple scales} (MMS) --- see e.g. \citet{bo78} for a general introduction. The MMS proceeds by assuming that all relevant quantities --- like the DF, potential etc --- are functions of two time variables, $t$ and $\tau = \varepsilon t$, \emph{whose variations are treated as independent}. 
This is an artifice whose efficacy is proved when, on obtaining an MMS solution, we set $\tau = \varepsilon t\,$ to get the desired solution to the full CBE eqn.(\ref{cbe3}); thus we write $f=f(\scrd, t, \tau)\,$, $\,\varphi = \varphi(\scrd, t, \tau)\,$, $\,\bfA_\bullet = \bfA_\bullet(t, \tau)\,$.
Then, replacing the time derivative $\,(\p/\p t)\,$ in the full CBE of eqn.(\ref{cbe3}) by $\,\left(\p/\p t + \varepsilon\,\p/\p\tau\right)\,$ we obtain the \emph{MMS--CBE}:
\beq
\frac{\p f}{\p t} \;+\; \varepsilon\frac{\p f}{\p \tau} \;+\; 
\Omega_{\rm k}\frac{\p f}{\p w} \;+\;
\varepsilon\left[\,f\,,\,\varphi \,+\, \bfr\cendot\bfA_\bullet\,\right]_{(6)}
\;=\; 0\,.
\label{cbemts}
\eeq

The next step in the MMS is to seek a solution to eqn.(\ref{cbemts}) that describes secular evolution plus small fluctuations. Then the dominant 
part of the DF must be independent of the fast time $t$, as well as the fast 
orbital phase $w$. Let us now gather the remaining 5 Delaunay variables and call them $\scrr\equiv\{I, L, L_z; g, h\}\,$. A point in 5--dim $\scrr$--space represents a Keplerian ellipse of given semi--major axis, eccentricity, inclination, periapse angle and nodal longitude. Henceforth such an ellipse will be referred to as a \emph{Gaussian Ring} or simply \emph{Ring}. We want to imagine secular evolution of $N\gg 1$ stars as the evolution of $N\gg 1$ Gaussian Rings over times $\sim\ts\,$. Hence, to O(1), the DF we seek must be a function only of the variables $(\scrr, \tau)$. With these considerations we begin with the following \emph{ansatz} for the DF, as a Fourier expansion in $w$\,:
\beq
f(\scrd, t, \tau) \;=\; \frac{1}{2\pi}\,F(\scrr, \tau) \;+\;
\frac{\varepsilon}{2\pi}\sum_{n=-\infty}^{\infty}f_n(\scrr, t, \tau)\exp{[{\rm i}n w]}
\;+\; O(\varepsilon^2)\,, 
\label{dffou}
\eeq
where $\,F\geq 0\,$ is the $O(1)$ \emph{secular DF}, and the 
$O(\varepsilon)$ fluctuation is given by its Fourier--coefficients $\,f_{-n} = f_n^\star\,$. 

We will derive equations for $F$ and all the $f_n$  by substituting the expansion of eqn.(\ref{dffou}) in the MMS--CBE eqn.(\ref{cbemts}), and ignoring $O(\varepsilon^2)$ terms. To do this we take the following preparatory steps:
\begin{itemize}
\item[{\bf (a)}]
The quantities, $\varphi(\scrd, t, \tau)$ and $\bfA_\bullet(t, \tau)$, are needed only to $O(1)$, which implies that we can set $f \to (2\pi)^{-1}\,F(\scrr, \tau)$ in eqns.(\ref{phiadel}). 

\item[{\bf (b)}]
To $O(1)$ the self--consistent potential is independent of $t$, and is given by:
\beq
\varphi(\scrd, \tau) \;=\; -\,GM_\bullet\int F(\scrr', \tau)\,{\rm d}\scrr'
\oint \frac{{\rm d}w'}{2\pi}\frac{1}
{\,\vert\bfr - \bfr'\vert\, }\,,
\label{phiF}
\eeq
where on the left side we have dropped the dependence on the fast time.
Fourier--expanding the right side in $w$, the potential can be written as:
\beq
\varphi(\scrd, \tau) \;=\;
\Phi(\scrr, \tau) \;+\; \sum_{n\neq 0}\varphi_n(\scrr, \tau)\exp{[{\rm i}n w]}\,,
\label{phifou}
\eeq
where
\begin{subequations}  
\begin{eqnarray}
\Phi(\scrr, \tau) &\;=\;& \int {\rm d}\scrr'\,F(\scrr', \tau)\,\Psi(\scrr, \scrr')\,;\label{phislow}\\[1ex]
\varphi_n(\scrr, \tau) &\;=\;& \int {\rm d}\scrr'\,F(\scrr', \tau)\,\psi_n(\scrr, \scrr')\,,
\label{phin}
\end{eqnarray}
\end{subequations}

\noindent
with the Ring--Ring interaction potential functions given by
\begin{subequations}
\begin{eqnarray}
\Psi(\scrr, \scrr') &\;=\;& -GM_\bullet\oint\oint\frac{{\rm d}w}{2\pi}\,
\frac{{\rm d}w'}{2\pi}\,\frac{1}{\left|\bfr - \bfr'\right|}\,,
\label{Psidef}\\[1em]
\psi_n(\scrr, \scrr') &\;=\;& -GM_\bullet\oint\oint\frac{{\rm d}w}{2\pi}\,
\frac{{\rm d}w'}{2\pi}\,\frac{\exp{[-{\rm i}n w]}}{\left|\bfr - \bfr'\right|}\,,\qquad n\neq 0\,.
\label{psindef}
\end{eqnarray}
\end{subequations}
Note that $\,\varphi_{-n} = \varphi_n^\star\,$ and $\,\psi_{-n} = \psi_n^\star\,$, because $\Phi$ and $\Psi$ are real quantities.

\item[{\bf (c)}] 
Using ${\rm d}\scrd = {\rm d}\scrr\,{\rm d}w\,$ in the second 
of eqns.(\ref{phiadel}), we get:
\beq
\bfA_\bullet \;=\; GM_\bullet\int
F(\scrr', \tau)\,{\rm d}\scrr'
\oint \frac{{\rm d}w}{2\pi}\,\frac{\hat{\bfr}\;}{r^2}
\;=\; {\bf 0}\,,
\label{azero}
\eeq
because $\oint {\rm d}w\,\hat{\bfr}/r^2 = {\bf 0}\,$, by the conservation of angular momentum along a Kepler orbit. So we can drop the $\bfr\cendot\bfA_\bullet$ term in the PB of eqn.(\ref{cbemts}).

\item[{\bf (d)}] 
The 6--dim PB of two functions, $\chi_1(\scrd)$ and $\chi_2(\scrd)$, can be expanded as:
\beq
\left[\,\chi_1\,,\,\chi_2\,\right]_{(6)} \;=\;
\left(\frac{\p \chi_1}{\p w}\frac{\p \chi_2}{\p I} -
\frac{\p \chi_1}{\p I}\frac{\p \chi_2}{\p w}\right) \;+\;
\left[\,\chi_1\,,\,\chi_2\,\right]\,,
\label{pb624}
\eeq
where the PB on the right side (without subscript),
\beq
\left[\,\chi_1\,,\,\chi_2\,\right] \;=\; 
\left(\frac{\p \chi_1}{\p g}\frac{\p \chi_2}{\p L} -
\frac{\p \chi_1}{\p L}\frac{\p \chi_2}{\p g}\right) \,+\, 
\left(\frac{\p \chi_1}{\p h}\frac{\p \chi_2}{\p L_z} -
\frac{\p \chi_1}{\p L_z}\frac{\p \chi_2}{\p h}\right)\,,
\label{pbdel4}
\eeq
is the 4--dim PB, whose action is restricted to the 4 dimensional
$I=\mbox{constant}\;$ surfaces in the 5 dimensional Ring--space.
In particular, for two functions $\chi_1(\scrr)$ and $\chi_2(\scrr)\,$, the 6--dim PB equals the 4--dim PB: $\;\left[\,\chi_1\,,\,\chi_2\,\right]_{(6)} \,=\, \left[\,\chi_1\,,\,\chi_2\,\right]\,$.
\end{itemize}

When eqns.(\ref{dffou}) and (\ref{phifou}) are substituted in the MMS--CBE 
eqn.(\ref{cbemts}), all the $O(1)$ terms are zero. Collecting the 
$O(\varepsilon)$ terms we get:
\begin{subequations}
\begin{eqnarray}
&&\frac{\p f_0}{\p t} \;+\; \frac{\p F}{\p \tau} \;+\; \left[\,F\,,\,\Phi\,\right] \;=\; 0\,;
\label{fluc0}\\[1em]
&&\frac{\p f_n}{\p t} \;+\; {\rm i}n\Omega_{\rm k}\,f_n \;=\; 
{\rm i}n\left(\frac{\p F}{\p I}\varphi_n \;+\; \frac{{\rm i}}{n}
\left[\,F\,,\,\varphi_n\,\right]\right)\,,\qquad\quad n\neq 0\,.
\label{flucn}
\end{eqnarray}
\end{subequations}

\noindent
The second and third terms on the left side of eqn.(\ref{fluc0}) are 
independent of the fast time $t$, so it is necessary that $\p f_0/\p t$
also be independent of $t$. Then the general solution is $f_0(\scrr, t, \tau) = \lambda_0(\scrr, \tau) + b(\scrr, \tau)\,t$, where $\lambda_0$ and 
$b$ are arbitrary real functions of $\scrr$ and $\tau$. However, 
$\vert f_0\vert$ is an increasing function of $t$ when $b \neq 0$. A cornerstone of the MMS is the requirement that physically admissible solutions must not grow on the fast time scale. This implies that $b=0$, 
and the MMS solution of eqn.(\ref{fluc0}) is:
\begin{subequations}
\begin{eqnarray}
\frac{\p F}{\p \tau} \;+\; \left[\,F\,,\,\Phi\,\right] \;=\; 0\,,
\label{cbesec}\\[1em]
f_0(\scrr, t, \tau) \;=\; \lambda_0(\scrr, \tau)\,.
\label{f0soln} 
\end{eqnarray}
\end{subequations}

\noindent
Eqn.(\ref{cbesec}) is the desired secular CBE for $F(\scrr, \tau)$, and is studied in more detail in the next two sections. From eqn.(\ref{f0soln}) we see that our $O(\varepsilon)$ MMS theory only requires $f_0$ to independent of $t$,  while letting it be an arbitrary real function of $\scrr$ and $\tau$. The right side of eqn.(\ref{flucn}) is independent of $t$, and it is straightforward to  write down the general solution:
\beq
f_n(\scrr, t, \tau) \;=\; \frac{1}{\Omega_{\rm k}}\left(\frac{\p F}{\p I}\varphi_n \;+\; \frac{{\rm i}}{n}\left[\,F\,,\,\varphi_n\,\right]\right) \;+\; \lambda_n(\scrr, \tau)\,\exp{[-{\rm i}n\Omega_{\rm k} t]} 
\,,\qquad\quad n\neq 0\,,
\label{fnsoln}
\eeq
\noindent
where the $\lambda_n$ are arbitrary functions of $\scrr$ and $\tau$, with 
$\lambda_{-n} = \lambda_n^\star\,$. Substituting eqns.(\ref{f0soln}) and (\ref{fnsoln}) in eqn.(\ref{dffou}), the general solution to the MMS--CBE eqn.(\ref{cbemts}) can be written as:
\begin{eqnarray}
f(\scrd, t, \tau) &\;=\;& \frac{1}{2\pi}F(\scrr, \tau) \;+\; 
\frac{\varepsilon}{2\pi\Omega_{\rm k}}\,\sum_{n\neq 0}\left(\frac{\p F}{\p I}\varphi_n \;+\; \frac{{\rm i}}{n}
\left[\,F\,,\,\varphi_n\,\right]\right)\exp{[{\rm i}n w]} 
\nonumber\\[1ex]
&&\qquad \;+\;
\frac{\varepsilon}{2\pi}\,\sum_{n=-\infty}^{\infty}\lambda_n(\scrr, \tau) 
\exp{[{\rm i}n (w - \Omega_{\rm k} t)]}
\;+\; O(\varepsilon^2)\,. 
\label{mmssoln}
\end{eqnarray}

Define two $t$--independent functions, $\Delta F(\scrd, \tau)$ and $\Lambda(\scrr, w, \tau)$, by the following Fourier series:
\begin{subequations}
\begin{eqnarray}
\Delta F(\scrd, \tau) &\;=\;& \frac{1}{\Omega_{\rm k}}\,\sum_{n\neq 0}\left(\frac{\p F}{\p I}\varphi_n \;+\; \frac{{\rm i}}{n}
\left[\,F\,,\,\varphi_n\,\right]\right)\exp{[{\rm i}n w]}\,,
\label{deltf}\\[1ex] 
\Lambda(\scrr, w, \tau) &\;=\;& 
\sum_{n=-\infty}^{\infty}\,\lambda_n(\scrr, \tau) 
\exp{[{\rm i}nw]}\,.
\label{fnxi}
\end{eqnarray}
\end{subequations}
The Fourier coefficients on the right side of eqn.(\ref{deltf}) are completely determined by $F\,$; they are well--defined because 
$n$ and $\Omega_{\rm k}$ are both non zero. We do not attempt to prove 
convergence of the Fourier series, and assume that this can be checked 
in any particular case. In eqn.(\ref{fnxi}) the $\lambda_n$ are arbitrary 
functions of $\scrr$ and $\tau$; they can be chosen such 
that the Fourier series converges and $\Lambda(\scrr, w, \tau)$ is 
a well-defined function of $\scrd$ and $\tau$. 
Having obtained the MMS solution, we go back to using a single 
time variable $t$, because we want a solution to the original problem of
the full CBE eqn.(\ref{cbe3}). In other words, $\tau$ is no longer regarded as an independent variable, but is simply $\tau = \varepsilon t$, 
which is a function of the true time variable $t$, and the small 
parameter $\varepsilon$. Setting $\tau = \varepsilon t$ in eqn.(\ref{mmssoln}), we obtain  
\beq
f(\scrd, t) \;=\; \frac{1}{2\pi}F(\scrr, \varepsilon t) \;+\;
\frac{\varepsilon}{2\pi}\Delta F(\scrd, \varepsilon t) \;+\;
\frac{\varepsilon}{2\pi}\Lambda\left(\scrr, \,w- \Omega_{\rm k} t, \,\varepsilon t\right)
\;+\; O(\varepsilon^2)\,. 
\label{fullsoln}
\eeq
\noindent
which is a solution of the full CBE eqn.(\ref{cbe3}), accurate to $O(\varepsilon)$ over times $\sim\ts$; this can be verified by direct substitution of eqn.(\ref{fullsoln}) in eqn.(\ref{cbe3}).

The DF of eqn.(\ref{fullsoln}) is the main result of this section. It  
is the sum of three terms: the first is $O(1)$ and proportional to $F\,$; 
the two other terms are $O(\varepsilon)$, and can be thought of as the small corrections to $F$. Some salient points to be noted are:
\begin{itemize} 
\item[{\bf 1.}] The secular DF $F$ is $O(1)$, independent of the orbital phase $w$, and evolves over times $\ts\,$, as determined by the CBE eqn.(\ref{cbesec}). 

\item[{\bf 2.}] The first $O(\varepsilon)$ correction is proportional to 
$\Delta F$ which is a fluctuating function of the orbital phase $w$ with zero mean. It is completely determined by $F$ and, like $F$, evolves over times $\ts\,$.  

\item[{\bf 3.}] The second $O(\varepsilon)$ correction is proportional to 
$\Lambda$ which is a function of $\,\scrr$, $\,(w-\Omega_{\rm k} t)\,$ and 
$\,\varepsilon t\,$.  Its functional form is undetermined in our 
$O(\varepsilon)$ MMS theory. However, it is important to note that the fast time $t$ occurs only in the form $(w-\Omega_{\rm k} t)$, which contributes only in a periodic manner, with no growth or decay.
\end{itemize}

The derivation of the full DF of eqn.(\ref{fullsoln}) given above makes it  
natural to think of the orbits followed by individual stars, to $O(1)$ accuracy, as slowly deforming Gaussian Rings. However, it is also useful to see this directly from stellar orbital dynamics. This is done in Appendix~A using the standard apparatus of secular perturbation theory. There it is proved: \emph{Every stellar orbit can be described to $O(1)$ accuracy as a Gaussian Ring whose dynamics is governed by the Hamiltonian, 
$\Phi(\scrr, \tau)$. Superimposed on this are $O(\varepsilon)$ oscillations on the fast time $\tk$, so we may think of a stellar orbit as a slowly evolving, `noisy' Gaussian Ring}.

MMS analysis is very useful because it deals directly with the DF --- bypassing consideration of individual stellar orbits --- and offers 
the following: for a given secular evolution described by the $O(1)$
DF $F(\scrr, \tau)$ in 5--dim $\scrr$ space, there is a full DF 
$f(\scrd, t)$ in the 6--dim $\scrd$ space for which the $O(\varepsilon)$ noise contributed by all the individual stellar orbits \emph{does not} act cooperatively to give rise to growth on the fast time scale $\tk$. 
\emph{Hence the fluctuations remain $O(\varepsilon)$ over the long secular time $\ts\,$, and the $O(1)$ dynamics described by $F$ is well--defined}. 
Thus MMS provides a consistent framework for secular dynamics, but it
would be well to note that this very consistency has been achieved 
at a certain cost. MMS filters out all $O(1)$ changes over fast orbital times $\tk$, and hence the resulting description is insensitive to either fast external perturbations or instabilities. It is generally expected that
these fast phenomena would have a period of growth followed by saturation. 
Further evolution --- including any instability --- unfolds over times $\sim\ts$. This is the regime that is accurately described by the MMS analysis of this Section.\footnote{An example of a fast instability is the well--known axisymmetric instability \citep{t64} applied to a Keplerian disc. All but the most unrealistically cold Keplerian discs would be stable by the Toomre criterion. But, as a matter of principle, we can indeed consider a Keplerian disc cold enough to be unstable. Then a secular description would make sense only after the instability has grown and saturated over times $\tk$, leaving the disc in a warmer state that is stable to all fast modes.}

\section{Collisionless Boltzmann equation for Gaussian Rings}

Henceforth we will study the $O(1)$ slow evolution described by the secular DF $F(\scrr, \tau)$. To this order the normalization of the full DF eqn.(\ref{norm}) implies that 
\beq
\int F(\scrr, \tau)\,{\rm d}\scrr \;=\; 1, 
\label{Fnorm}
\eeq 
so we may regard $F(\scrr, \tau)$ as a probability distribution function 
in 5--dim Ring space. We shall refer to $F$ as the \emph{Ring distribution 
function}. The time evolution of $F$ is described by the secular CBE 
eqn.(\ref{cbesec}), $\;(\p F/\p \tau) + [F, \Phi] = 0\,$. We recall that 
$\Phi(\scrr, \tau)$ is the (scaled) orbit--averaged self--gravitational potential of the stars which is determined self--consistently by $F$ itself:
\beq
\Phi(\scrr, \tau) \;=\; \int {\rm d}\scrr'\,F(\scrr', \tau)\,\Psi(\scrr, \scrr')\,,
\label{selfgrav}
\eeq
\noindent
where 
\beq
\Psi(\scrr, \scrr') \;=\; -GM_\bullet\oint\oint\frac{{\rm d}w}{2\pi}\,
\frac{{\rm d}w'}{2\pi}\,\frac{1}{\left|\bfr - \bfr'\right|}
\label{rrint}
\eeq  
\noindent
is the (scaled) Ring--Ring interaction potential energy. Note that $\Phi$
is independent of $\varepsilon$, so the secular evolution of eqn.(\ref{cbesec}) has the property of \emph{mass--scale invariance}: i.e. the behaviour of $F$ as a function of $\scrr$ and $\tau$ is independent of 
$\varepsilon$ when $\varepsilon \ll 1\,$. This means that, for given MBH mass $M_\bullet$, the evolution of the stellar system appears independent of its mass $M$, so long as $M \ll M_\bullet\,$ and the time variable used is $\tau = \varepsilon t$. This invariance holds when every star is acted upon 
by the Keplerian potentials of the MBH and all the other stars. We now add two corrections to $\Phi\,$; these are the orbit--averaged effects of (a) relativistic corrections and (b) external perturbations.

\noindent
{\bf (a) \emph{Relativistic Corrections}} to the gravity of the MBH
introduce post--Newtonian (PN) corrections, the most important ones being the 1~PN Schwarzschild precession of the apses, and the 1.5~PN 
Lense--Thirring precession of the apses and nodes for a spinning MBH. Both 
effects are described by the relativistic (secular) Hamiltonian $\,\varepsilon H^{\rm rel}\,$:

\begin{eqnarray}
H^{\rm rel}(I, L, L_z) &\;=\;&  H^{PN1}(I, L) \;+\; H^{PN1.5}(I, L, L_z)\,,
\nonumber\\[3ex]
H^{PN1}(I, L) &\;=\;& -\,B_1\,\frac{1}{I^3 L}\qquad\mbox{with}\qquad
B_1 \;=\; \frac{3(GM_\bullet)^4}{c^2}\,\frac{M_\bullet}{M}\,,
 \nonumber\\[3ex]
H^{PN1.5}(I, L, L_z) &\;=\;& B_{1.5}\,\frac{L_z}{I^3 L^3}\,,\qquad\mbox{with}\qquad
B_{1.5} \;=\; \frac{2(GM_\bullet)^5}{c^3}\,\frac{M_\bullet}{M}\,\sigma\,,
\label{hrel}
\end{eqnarray}
\noindent
We have assumed that the spin of the MBH points along the $z$--axis; 
$\,0\leq \sigma\leq 1\,$ is the spin parameter. Note that the definitions of the constants $B_1$ and $B_{1.5}\,$ include the factor 
$(M_\bullet/M) = \varepsilon^{-1}$.

\noindent 
{\bf (b) \emph{External Perturbations}} on the stars as well as the MBH --- due to nuclear density cusps and/or slowly moving distant masses --- can also be included in the secular description. Let $\,\varepsilon\Phi^{\rm ext}(\scrr, \tau)\,$ be the orbit--averaged contribution and $\,\varepsilon\bfA^{\rm ext}(\tau)$ be the acceleration of the MBH. These will contribute an orbit--averaged tidal potential $\,\varepsilon\Phi^{\rm tid}(\scrr, \tau)\,$ to the secular Hamiltonian for a star, where  
\beq
\Phi^{\rm tid}(\scrr, \tau) \;=\; \Phi^{\rm ext}(\scrr, \tau) \;+\; \bfX(\scrr)\cdot\bfA^{\rm ext}_\bullet(\tau)\,,
\label{tidpot}
\eeq
\noindent
with
\beq
\bfX(\scrr) \;=\; \oint \frac{{\rm d}w}{2\pi}\,\bfr(\scrd)
\eeq
\noindent
equal to the orbit--averaged position vector. Note that, similar to the case 
of relativistic perturbations, $\,\Phi^{\rm tid}\,$ has a factor 
$\varepsilon^{-1}$ implicit in its definition.

Putting the pieces together, the general Hamiltonian for a Gaussian Ring 
is:
\beq
H(\scrr, \tau) \;=\; \Phi(\scrr, \tau) \;+\; H^{\rm rel}(I, L, L_z) \;+\; \Phi^{\rm tid}(\scrr, \tau)\,,
\label{secham}
\eeq
\noindent
with the Ring equations of motion:
\begin{eqnarray}
I &\;=\;& \sqrt{GM_\bullet a} \;=\; \mbox{constant}\,,
\nonumber\\[1em]
\frac{{\rm d}L}{{\rm d}\tau} &\;=\;& -\,\frac{\p H}{\p g}\,,\qquad  
\frac{{\rm d}g}{{\rm d}\tau} \;=\; \frac{\p H}{\p L}\,;\qquad
\frac{{\rm d}L_z}{{\rm d}\tau} \;=\; -\,\frac{\p H}{\p h}\,,\qquad  
\frac{{\rm d}h}{{\rm d}\tau} \;=\; \frac{\p H}{\p L_z}\,.
\label{eom}
\end{eqnarray}
\noindent
A secular \emph{integral of motion} is a function $\,\scri(\scrr, \tau)$ that is constant along the orbit of a Ring:\footnote{Henceforth we drop 
the word `secular' when referring to an integral of motion.} 
\beq
\frac{{\rm d}}{{\rm d}\tau}\scri\left(\scrr(\tau), \tau\right) \;=\; 0\,.
\eeq
\noindent
\emph{It is a feature of secular dynamics that $I=\sqrt{GM_\bullet a}\,$ is an integral of motion for quite general and time--dependent 
$H(\scrr, \tau)$. When the Hamiltonian is time--independent, it provides a second integral $H(\scrr)$}. 

The equation governing the evolution of the distribution of Gaussian Rings is obtained by replacing $\Phi$ by $H$ in eqn.(\ref{cbesec}):
\beq
\frac{\p F}{\p \tau} \;+\; \left[\,F\,,\,H\,\right] \;=\;0\,.
\label{cbe-des}
\eeq
\noindent
The time evolution is collisionless: using eqns.(\ref{eom}) and (\ref{cbe-des}) we can verify that 
\begin{eqnarray}
\frac{{\rm d}F}{{\rm d}\tau} &\;\equiv\;& 
\frac{\p F}{\p t} \;+\; \frac{{\rm d}L}{{\rm d} \tau}\frac{\p F}{\p L} \;+\; 
\frac{{\rm d}g}{{\rm d} \tau}\frac{\p F}{\p g} \;+\; 
\frac{{\rm d}L_z}{{\rm d} \tau}\frac{\p F}{\p L_z} \;+\; 
\frac{{\rm d}h}{{\rm d} \tau}\frac{\p F}{\p h}
\nonumber\\[1em]
&\;=\;&
\frac{\p F}{\p \tau} \;-\; \frac{\p H}{\p g}\frac{\p F}{\p L} \;+\; \frac{\p H}{\p L}\frac{\p F}{\p g}
\;-\; \frac{\p H}{\p h}\frac{\p F}{\p L_z} \;+\; 
\frac{\p H}{\p L_z}\frac{\p F}{\p h}\nonumber\\[1em]
&\;=\;& \frac{\p F}{\p \tau} \;+\; \left[\,F\,,\,H\,\right] \;=\; 0\,,
\label{cbetwo}
\end{eqnarray}
\noindent
Therefore eqn.(\ref{cbe-des}) describes a system of Gaussian Rings evolving collisionlessly under the combined actions of self--gravity, relativistic corrections to the MBH's Newtonian gravity and slow external perturbations. It will be referred to as the \emph{Ring CBE}. 

The Ring Hamiltonian $H$ of eqn.(\ref{secham}) is in general not independent of $\varepsilon$, because both $H^{\rm rel}$ and $\Phi^{\rm tid}$ are proportional to $\varepsilon^{-1}$. Therefore the property of mass--scale invariance of the Ring CBE is in general broken. Another important general property, valid for quite general $F(\scrr, \tau)$ and $H(\scrr, \tau)$, is this: \emph{The probability for a Ring to be in $(I,\, I + {\rm d}I)$ is a conserved quantity}. In other words the PDF in 1--dim $I$--space, defined by
\beq
P(I) \;=\; \int\, dL\,dL_z\,dg\,dh\, F(I,\, L,\, L_z,\, g,\, h,\,\tau)\,, 
\label{conserved}
\eeq
\noindent
is independent of $\tau$, as can be verified directly using the Ring CBE eqn.(\ref{cbe-des}). The physical reason for this is that the semi--major axis of every Ring is conserved, so the function $P(I)$ must stay frozen as $F(\scrr, \tau)$ evolves over secular times. Therefore secular evolution is driven by gravitational torques between Gaussian Rings, with no exchange of Keplerian energies.

\subsection{Collisionless Boltzmann equation for Zero--thickness Flat Discs}

It is useful to study the idealized case when all the Rings are
confined to the $xy$ plane. In these zero--thickness discs, the angular momentum of every Ring points along $\pm\hat{z}$, so that
a Ring is specified by its semi--major axis, angular momentum and apsidal 
longitude. The Ring phase space is now 3--dim: we write $\scrr = \{I, L, g\}$, where $L$ stands for the angular momentum which can now take both positive and negative values $\left(-I\leq L\leq I\,\right)$, and $g$ is the longitude of the periapse.\footnote{This is the convention for the symbols 
$(L, g)$ we will always use when dealing with Ring dynamics that is confined to the $xy$ plane.} Rings with $L >0$ will be referred to as \emph{prograde} and Rings with $L <0$ will be referred to as \emph{retrograde}. 

The general time--dependent Ring DF is $F(I, L, g, \tau)$. The (scaled)
orbit--averaged self--gravitational potential, $\Phi(I, L, g, \tau)$,
is given in terms of $F$ and $\Psi(\scrr, \scrr')$ by eqns.(\ref{selfgrav}) and (\ref{rrint}). To calculate $\Psi(\scrr, \scrr')$, we express $\bfr=(x, y)$ in terms of the 2--dim Delaunay variables:
\begin{eqnarray}
\qquad\qquad x &\;=\;& a(\cos\eta - e)\cos g \,\;\mp\;\, a\sqrt{1-e^2\,}\sin\eta\,\sin g\,,
\nonumber\\[1ex]
y &\;=\;& a(\cos\eta - e)\sin g \,\;\pm\;\, a\sqrt{1-e^2\,}\sin\eta\,\cos g\,,
\label{xydel}
\end{eqnarray}
where the upper(lower) signs in the second terms on the right hand side
correspond to prograde(retrograde) orbits. As earlier, $\eta$ is related 
to the mean anomaly by $w = \eta - e\sin\eta$. Using corresponding expressions for $\bfr'=(x', y')$, we see that $\left|\bfr - \bfr'\right|^{-1}$ is an explicitly known function of the Delaunay variables, $\scrd$ and $\scrd'$, and can be averaged over $w$ and $w'$.  Then the secular Hamiltonian governing Ring dynamics is:
\beq
H(I, L, g, \tau) \;=\; \Phi(I, L, g, \tau) \;+\; H^{\rm rel}(I, L) \;+\; \Phi^{\rm tid}(I, L, g, \tau)\,,
\label{secham-rt}
\eeq
\noindent
where the relativistic correction,
\beq
H^{\rm rel}(I, L) \;=\; -B_1\frac{1}{I^3\vert L\vert} \;+\;
B_{1.5} \frac{{\rm Sgn}(L)}{I^3L^2}\,,
\label{hrel-rt}
\eeq
\noindent
takes a slightly different form from eqn.(\ref{hrel}), because we have
allowed $L$ to take both positive and negative values for zero--thickness flat discs. The Ring equations of motion are:
\beq
I \;=\; \sqrt{GM_\bullet a} \;=\; \mbox{constant}\,,\qquad\quad
\frac{{\rm d}L}{{\rm d}\tau} \;=\; -\,\frac{\p H}{\p g}\,,\qquad\quad  
\frac{{\rm d}g}{{\rm d}\tau} \;=\; \frac{\p H}{\p L}\,.
\label{eom-rt}
\eeq
\noindent
The DF satisfies the Ring CBE,
\beq
\frac{{\rm d}F}{{\rm d}\tau} \;=\; 
\frac{\p F}{\p \tau} \;-\; \frac{\p H}{\p g}\frac{\p F}{\p L} \;+\; \frac{\p H}{\p L}\frac{\p F}{\p g} \;=\; 
\frac{\p F}{\p \tau} \;+\; \left[\,F\,,\,H\,\right] \;=\; 0\,,
\label{cbe-rt}
\eeq
\noindent
where the Poisson Bracket is now 2--dim, because it involves only the 
canonically conjugate pair $(L, g)$. As earlier, the probability for a Ring to be in $(I,\, I + {\rm d}I)$ is a conserved quantity: i.e. the PDF in 1--dim $I$--space, defined by
\beq
P(I) \;=\; \int\, dL\,dg\,\, F(I,\, L,\, g, \,\tau)\,, 
\label{conserved-rt}
\eeq
\noindent
is independent of $\tau$, as can be verified directly using the Ring CBE 
eqn.(\ref{cbe-rt}).

\section{Secular Collisionless Equilibria}

Secular collisionless equilibria are described by DFs $F$ that are independent of $\tau$; these DFs will also be referred to as \emph{stationary}. From eqn.(\ref{cbe-des}) a stationary DF must satisfy $\left[F, H\right] = 0\,$. These can be constructed using a secular version of Jeans theorem. 

\medskip 
\noindent
{\bf \emph{Secular Jeans theorem}:} Any stationary solution of the Ring CBE eqn.(\ref{cbe-des}) is a function of $\scrr$ only through the time--independent integrals of motion of $H(\scrr)$, and any function of the time--independent integrals of $H(\scrr)$ is a stationary solution of eqn.(\ref{cbe-des}).

{\bf \emph{Proof}:} This is similar to the one given in \citet{bt08}:
If $F$ is a stationary solution of eqn.(\ref{cbe-des}), then it is also an integral of motion; so the first part of the theorem is proved. Conversely, if $\;\scri_1(\scrr),\ldots, \scri_n(\scrr)\,$ be $n$ time--independent integrals, then any $F(\scri_1,\ldots,\scri_n)$ satisfies the Ring CBE:
\beq
\frac{{\rm d}}{{\rm d}\tau} F\left(\scri_1(\scrr),\ldots,\scri_n(\scrr)\right)
\;=\; \sum_{m=1}^n \frac{\p F}{\p \scri_m}
\frac{{\rm d}\scri_m}{{\rm d}\tau} \;=\; 0\,,
\eeq
\noindent
because $({\rm d}\scri_m/{\rm d} \tau) = 0$ for all $m=1,\ldots,n\,.\quad\square$

\medskip
\noindent
In practice a stronger version of the Secular Jeans theorem is used, by 
requiring a stationary DF to be a function of the \emph{isolating
integrals of motion}. In contrast with non--secular dynamics two isolating
integrals, $I$ and $H(\scrr)$, are always available in secular dynamics. 
Additional symmetries can promote more isolating integrals. Hence a richer variety of stationary DFs can be constructed, with greater ease than in non secular dynamics. Below we discuss briefly the simplest cases. 

\subsection{Spherical Equilibria}

For a spherically symmetric stellar system around a MBH, the
orbit--averaged self--gravitational potential $\Phi$ is a function only
of $I$ and $L$. An external potential can be included if it is also
spherically symmetric with respect to the MBH; this causes no acceleration 
of the MBH, so $\Phi^{\rm tid} =\Phi^{\rm ext}$ are also functions only of $I$ and $L$. Therefore the Ring Hamiltonian of 
eqn.(\ref{secham}),
\beq
H(I, L, L_z) \;=\; \Phi(I, L) \;+\; H^{\rm rel}(I, L, L_z) \;+\; \Phi^{\rm ext}(I, L)\,,
\label{hamsph}
\eeq
\noindent
is independent of the angle variables. Using eqn.(\ref{eom}) the orbit of a Ring is given by:
\begin{eqnarray}
I &\;=\;& \mbox{constant}\,,\qquad L \;=\; \mbox{constant}\,,
\qquad L_z \;=\; \mbox{constant}\,,
\nonumber\\[1em]
\frac{{\rm d}g}{{\rm d}\tau} &\;\equiv\;& \Omega(I, L, L_z) \;=\; \frac{\p H}{\p L}\;; \qquad\mbox{apse precession frequency}\,,
\nonumber\\[1em]
\frac{{\rm d}h}{{\rm d}\tau} &\;\equiv\;& \nu(I, L)\;=\; \frac{\p H}{\p L_z} \;=\; \frac{B_{1.5}}{(IL)^3}\;; \qquad\mbox{node precession frequency}\,.
\label{orbsph}
\end{eqnarray}
\noindent
Therefore, every Ring in a spherical system has constant semi--major axis, 
eccentricity and inclination; its periapse precesses at a constant 
rate $\Omega(I, L, L_z)$; its ascending node precesses at a constant, 
inclination--independent rate $\nu(I, L)$, due solely to 1.5 PN relativistic 
effects for a spinning MBH. The Ring orbital dynamics is integrable, with the Delaunay actions $(I, L, L_z)$ serving as three isolating integrals of motion. 

The Secular Jeans theorem implies that a general stationary spherical DF
can be a function of $I$, $L$ and $L_z\,$. We will restrict attention to 
DFs which have complete spherical symmetry which these are of the form $F(I, L)$. It should be noted that the stellar distribution in phase space is given \emph{explicitly} to $O(1)$, without having to solve an integral equation. Indeed, the functional dependence on the six coordinates $\{\bfr, \bfu\}$ is obtained by setting
\beq 
I \;=\;  
GM_\bullet\left[\frac{2GM_\bullet}{r} \,-\, u^2\right]^{-1/2}\,,\qquad\qquad 
L \;=\; \vert\bfr\cross\bfu\vert\,,
\label{cart}
\eeq
\noindent
in the DF $F(I, L)$. Then, observationally interesting quantities --- such
as the radial profiles of the surface density, the velocity dispersion tensor and the line--of--sight velocity distribution (LOSVD) 
--- can be calculated by direct integration of the DF. It is readily seen that the DFs $F(I, L)$ describe systems with zero streaming motions. The velocity dispersion tensor is in general anisotropic.

\subsection{Axisymmetric Equilibria}

When the stellar system is axisymmetric about the MBH with symmetry
axis along $\hat{z}$, the orbit--averaged self--gravitational potential 
$\Phi$ is a function of ($I,\,L,\,L_z,\, g$). An external potential can be included if it is also axisymmetric with respect to the MBH with the 
same symmetry axis; this causes no acceleration of the MBH, so $\Phi^{\rm tid} =\Phi^{\rm ext}$ are also functions only of
($I,\,L,\,L_z,\, g$). Therefore the Ring Hamiltonian of eqn.(\ref{secham}) simplifies to
\beq
H(I, L, L_z, g) \;=\; \Phi(I, L, L_z, g) \;+\; H^{\rm rel}(I, L, L_z) \;+\; \Phi^{\rm ext}(I, L, L_z, g)\,.
\label{hamaxi}
\eeq
\noindent
Using eqn.(\ref{eom}), the orbit of a Ring is given by
\beq
I \;=\; \mbox{constant}\,,\qquad L_z \;=\; \mbox{constant}\,,\qquad
\frac{{\rm d}g}{{\rm d}\tau} \;=\; \frac{\p H}{\p L}\,,\qquad
\frac{{\rm d}h}{{\rm d}\tau} \;=\; \frac{\p H}{\p L_z}\,. 
\label{orbaxi}
\eeq
\noindent
As a Ring evolves with constant $I$ and $L_z$, its eccentricity, 
inclination, apse and node can all vary with time. From the Secular Jeans theorem, a general stationary axisymmetric DF must be of the form $F(I, L_z, H)$. Since $H = \Phi + H^{\rm rel} + \Phi^{\rm ext}$ is a function of $(I, L, L_z, g)$, so is the DF. However, to express the DF explicitly as a function of the phase space coordinates, we need to determine $\Phi(I, L, L_z, g)$ by solving the self--consistent problem of eqns.(\ref{selfgrav}) and (\ref{hamaxi}):
\beq
\Phi(I, L, L_z, g) \;=\; \int {\rm d}I'\,{\rm d}L'\,{\rm d}L_z'\,{\rm d}g'\,\,F(I', L'_z, H')\oint {\rm d}h'\,\Psi(\scrr, \scrr')\,,
\label{selfaxi}
\eeq
\noindent
where $H' = H(I', L', L_z', g')$.\footnote{The left side of eqn.(\ref{selfaxi}) indicates that $\Phi$ is independent of $h$, as must be the case for any axisymmetric system. We can check that the right side of eqn.(\ref{selfaxi}) is also independent of $h$: the nodal longitudes occur in $\Psi(\scrr, \scrr')$ only in the combination $(h - h')$. So, in the integral of $\Phi$ over $h'$, when we change the integration variable from $h'$ to
$(h' - h)$, we get a quantity that is independent of $h$.}

The problem simplifies considerably for DFs that are of the form $F(I, L_z)$. By replacing $I$ and $L_z$ by
\beq 
I \;=\;  
GM_\bullet\left[\frac{2GM_\bullet}{r} \,-\, u^2\right]^{-1/2}\,,\qquad\qquad 
L_z \;=\; \hat{z}\cendot\left(\bfr\cross\bfu\right)\,,
\label{cart2}
\eeq
\noindent
we see that, just like in the spherical case, the DF is known explicitly to $O(1)$ as a function of $\{\bfr, \bfu\}$, without having to solve an integral equation. Then observationally interesting quantities --- such as the sky--maps of the surface density, the velocity dispersion tensor and the line--of--sight velocity distribution (LOSVD) --- can be calculated by direct integration of the DF.

\bigskip
\noindent
{\bf (a) \emph{Zero--thickness Flat Discs}:} We recall from \S~4.1 that 
the Ring space is now 3--dim; $\scrr = (I, L, g)$, where $-I\leq L \leq I$ is the angular momentum and $g$ is the apsidal longitude.  For an axisymmetric disc around a MBH, the (scaled) orbit--averaged self--gravitational potential $\Phi$ is a function only of $I$ and $L$. An external potential can be included if it is also axisymmetric with respect to the MBH; this causes no acceleration of the MBH, so $\,\Phi^{\rm tid} =\Phi^{\rm ext}$ are also functions only of $I$ and $L$. Therefore the Ring Hamiltonian of eqn.(\ref{secham-rt}),
\beq
H(I, L) \;=\; \Phi(I, L) \;+\; H^{\rm rel}(I, L) \;+\; \Phi^{\rm ext}(I, L)\,,
\label{hamaxi2}
\eeq
\noindent
is independent of the angle variables. Using eqn.(\ref{eom-rt}) the orbit of a Ring is given by:
\beq
I \;=\; \mbox{constant}\,,\qquad L \;=\; \mbox{constant}\,,\qquad
\frac{{\rm d}g}{{\rm d}\tau} \;\equiv\; \Omega(I, L) \;=\; \frac{\p H}{\p L}\,. 
\label{orbaxi2}
\eeq
\noindent
Therefore, every Ring in an axisymmetric system has constant semi--major axis and eccentricity, with its apsidal longitude precessing at the constant rate $\Omega(I, L)$. The Ring orbital dynamics is integrable, with the Delaunay actions $(I, L)$ serving as two isolating integrals of motion. 
The Secular Jeans theorem implies that a general stationary axisymmetric 
DF has the form $F(I, L)$. The stellar distribution in phase space is given explicitly to $O(1)$, without having to solve a self--consistent problem. Indeed, the functional dependence on the four coordinates $\{x,y, u_x, u_y\}$ is obtained by setting
\beq 
I \;=\;  
GM_\bullet\left[\frac{2GM_\bullet}{r} \,-\, u^2\right]^{-1/2}\,,\qquad\qquad L \;=\; xu_y \,-\, yu_x\,,
\label{cart2new}
\eeq

\noindent
in the DF $F(I, L)$. Similar to 3--dim axisymmetric systems, $F(I, L_z)$
discussed above, observationally interesting quantities --- such as the velocity dispersion tensor or the LOSVD --- can be calculated by direct integration of the DF. Since $L$ can take both positive and negative values, a general $F(I, L)$ describes discs with or without rotation, and generally anisotropic velocity dispersions. These DFs are relevant to studies such as \citet{tt14}, who solved for maximum entropy equilibria of self--gravitating Keplerian disks (in the case where all particles have the same semi--major axis). When $F(I, L)$ is an even function of $L$, local streaming motions are absent and the disc is non--rotating with anisotropic velocity dispersions. A special case consists of DFs of the form $F(I)$, which are non--rotating and have isotropic velocity dispersions.

\section{Perturbations and Stability of Secular Equilibria}

If $F_0(\scrr)$ be a stationary DF, the Hamiltonian governing the orbits of Gaussian Rings in the unperturbed system is $H_0(\scrr)= \Phi_0(\scrr) + H^{\rm rel}(I, L, L_z) + \Phi_0^{\rm tid}(\scrr)$, and we must have $[F_0\,, \,H_0] = 0$.  Let $F_1(\scrr, \tau)$ be an infinitesimal perturbation about $F_0$. This gives rise to an infinitesimal perturbation of the self--gravitational potential, 
\beq 
\Phi_1(\scrr, \tau) \;=\; \int {\rm d}\scrr'\,F_1(\scrr', \tau)\,\Psi(\scrr, \scrr')\,.
\label{phipert}
\eeq
\noindent
Let the external potential also be allowed to change infinitesimally, 
so it contributes an additional $\Phi_1^{\rm tid}(\scrr, \tau)\,$. Then the infinitesimal change in the Ring Hamiltonian is 
\beq
H_1(\scrr, \tau) \;=\; \Phi_1(\scrr, \tau) \;+\; 
\Phi_1^{\rm tid}(\scrr, \tau)\,.
\label{ham-pert}
\eeq
\noindent
Substituting $F = F_0 + F_1$ and $H = H_0 + H_1$ in the Ring CBE eqn.(\ref{cbe-des}), and keeping only the linear terms, gives the \emph{linearized collisionless Boltzmann equation} (LCBE):
\beq
\frac{\p F_1}{\p \tau} \;+\; \left[\,F_1\,,\,H_0\,\right]
\;+\; \left[\,F_0\,,\,H_1\,\right] \;=\;0\,.
\label{lcbe}
\eeq
\noindent 
Eqns.(\ref{phipert})---(\ref{lcbe}) define the linear integral equation for the perturbation $F_1(\scrr, \tau)$. When $\Phi_1^{\rm tid} =0$, the evolution of an infinitesimal perturbation is determined by only its self--gravity $\Phi_1(\scrr, \tau)$ --- and the unperturbed flows due to $H_0(\scrr)$ in which we have allowed for an external contribution $\Phi_0^{\rm tid}(\scrr)$. The unperturbed DF $F_0(\scrr)$ is said to be \emph{secularly stable} if there is no growing solution $F_1$. Below we discuss two problems: 
\begin{itemize}
\item[{\bf (a)}] Long--lived, warp--like perturbations of spherically symmetric DFs of the form $F_0(I, L)$. 

\item[{\bf (b)}] Axisymmetric, zero--thickness, flat discs with DFs of the form $F_0(I, L)$. 
\end{itemize}

\subsection{Perturbations of Spherical Equilibria}

For a completely spherical unperturbed system the DF is $F_0(I, L)$, 
with the unperturbed Ring Hamiltonian given by eqn.(\ref{hamsph}):
\beq
H_0(I, L, L_z) \;=\; \Phi_0(I, L) \;+\; H^{\rm rel}(I, L, L_z) \;+\; 
\Phi_0^{\rm ext}(I, L)\,.
\label{hamsph0}
\eeq
\noindent
From eqn.(\ref{orbsph}), the apse precession frequency is $\Omega(I, L, L_z) = \p H_0/\p L$; and the inclination--independent node precession frequency, $\nu(I, L) = B_{1.5}(IL)^{-3}\,$, is due only to the 1.5 PN relativistic correction for a spinning MBH. Then the LCBE of eqn.(\ref{lcbe}) can be written as:
\beq
\frac{\p F_1}{\p \tau} \;+\; \Omega\frac{\p F_1}{\p g}
\;+\; \nu\frac{\p F_1}{\p h} \;=\; 
\frac{\p F_0}{\p L}\frac{\p H_1}{\p g}\,.
\label{lcbe-sph}
\eeq
\noindent
where $H_1 = \Phi_1 + \Phi_1^{{\rm tid}}$ contains the effects of the 
perturbed self--gravity and arbitrary external (slow) perturbation. 
From this equation for the linear (secular) perturbation $F_1(\scrr, \tau)$, we can arrive at the following general conclusion: \emph{Any growing or decaying behaviour of $F_1$ must necessarily be $g$--dependent, even 
in the presence of slow external perturbations of arbitrary form}. To see this let us define the $g$--averaged perturbed DF:
\beq
\bar{F}_1(I, L, L_z, h, \tau) \;=\; 
\oint\frac{{\rm d}g}{2\pi}\, F_1(I, L, L_z, g, h, \tau)\,, 
\eeq
which can be thought of as the DF for stellar orbits, each of which is smeared into an axisymmetric annulus whose inner/outer radii are
equal to the peri/apo centre distances. Averaging eqn.(\ref{lcbe-sph})
over $g$, we get 
\beq
\frac{\p \bar{F}_1}{\p \tau} \;+\; \nu(I, L)\frac{\p \bar{F}_1}{\p h} \;=\; 0\,.
\label{stab-sph-g-ave}
\eeq
The general solution is
\beq
\bar{F}_1(I, L, L_z, h, \tau) \;=\; Q(I, L, L_z, h - \nu\tau)\,,
\label{soln-warp} 
\eeq
where $Q(I, L, L_z, h)$ is an arbitrary initial $g$--independent perturbation. According to eqn.(\ref{soln-warp}) such an initial condition 
evolves through precession of the nodes at the rate $\nu(I, L)$ without either growing or decaying. The nodal precession around the spin axis of the 
MBH will be negligible in non--relativistic stellar systems. Then the perturbations are quasi--static, and could describe \emph{long--lived warps in spherical nuclear clusters}. Since the time evolution --- in the linear regime --- is independent of arbitrary external tidal fields $\Phi_1^{{\rm tid}}(\scrr, \tau)$, the perturbed DF retains memory of its origins. 

A linear perturbation must be $g$--dependent for it to either grow or decay.
We do not investigate these, and refer the reader to \citet{tre05,pps07}
which were discussed briefly in the Introduction. \emph{Spherical isotropic DFs} of the form $F_0(I)$ are special because the right hand side of eqn.(\ref{lcbe-sph}) vanishes. Then the LCBE is:
\beq
\frac{\p F_1}{\p \tau} \;+\; \Omega\frac{\p F_1}{\p g}
\;+\; \nu\frac{\p F_1}{\p h} \;=\; 0\,,
\label{lcbe-iso}
\eeq

\noindent
Therefore an arbitrary infinitesimal perturbation of an isotropic distribution function $F_0(I)$ flows collisionlessly along the unperturbed trajectories in $\scrr$--space.  The general solution of eqn.(\ref{lcbe-iso}) is:
\beq
F_1(\scrr, \tau) \;=\; Q(I, L, L_z, \,g - \Omega\tau, \,h - \nu\tau)\,,\qquad
\qquad\mbox{for $\,\tau \geq 0\,$.}
\label{iso-soln}
\eeq

\noindent
where $Q(\scrr) = Q(I, L, L_z, g, h)$ is an arbitrary initial perturbation that is assumed to be given. This implies: (a) All isotropic DFs $F_0(I)$ are secularly stable, because the perturbation evidently is non growing; 
(b) The evolution of an initial perturbation is independent of $H_1(\scrr, \tau)$.

\subsection{Stability of Axisymmetric Zero--thickness Flat Discs}

We recall from \S~4.1 that, for zero--thickness flat discs, the Ring phase space is 3--dim: $\scrr = (I, L, g)$, where $-I\leq L \leq I$ is the angular momentum, and $g$ is the apsidal longitude. Stationary, 
axisymmetric DFs are of the form $F_0(I, L)$, with the corresponding 
Hamiltonian of eqn.(\ref{hamaxi2}):
\beq
H_0(I, L) \;=\; \Phi_0(I, L) \;-\; B_1\frac{1}{I^3\vert L\vert} \;+\;
B_{1.5} \frac{{\rm Sgn}(L)}{I^3L^2} \;+\; \Phi_0^{\rm ext}(I, L)\,,
\label{hamaxi3}
\eeq

\noindent
From eqn.(\ref{orbaxi2}) we see that, in the unperturbed system, every Ring precesses rigidly at the apse precession rate $\Omega(I, L) = \p H_0/\p L\,$. A linear perturbation, $F_1(\scrr, \tau)$, obeys the LCBE:
\beq
\frac{\p F_1}{\p \tau} \;+\; \Omega\frac{\p F_1}{\p g}
\;=\; \frac{\p F_0}{\p L}\frac{\p H_1}{\p g}\,,
\label{lcbe-axi}
\eeq
where
\begin{subequations} 
\begin{eqnarray}
H_1(I, L, g, \tau) &\;=\;& \Phi_1(I, L, g, \tau) \;+\; \Phi_1^{\rm tid}(I, L, g, \tau)\,,
\label{ham1-axi}\\[1ex]
\Phi_1(I, L, g, \tau) &\;=\;& \int {\rm d}I'\,{\rm d}L'\,{\rm d}g'\,\,\Psi(I, L, g, I', L', g')\,F_1(I', L', g', \tau)\,.
\label{pert-axi}
\end{eqnarray}
\end{subequations}
The linear problem posed by eqns.(\ref{lcbe-axi}) and (\ref{pert-axi}) is still difficult to solve, because of the integral relationship between 
$\Phi_1$ and $F_1$ in eqn.(\ref{pert-axi}). 

Below we derive a stability result, by exploiting the symmetry properties of the Ring--Ring interaction potential function $\Psi(\scrr, \scrr')$. We write:
\beq
\Psi(I, L, g, I', L', g') \;=\; -GM_\bullet\oint\oint\frac{{\rm d}w}{2\pi}\,
\frac{{\rm d}w'}{2\pi}\,\frac{1}{\left|\bfr - \bfr'\right|}\,,
\label{rrint-disc}
\eeq  

\noindent
where $\bfr = (x, y)$ and $\bfr' = (x', y')$ can be written in terms 
of $(I, L, \eta, g)$ and $(I', L', \eta', g')$, respectively, using eqn.(\ref{xydel}); we also recall that ${\rm d}w = (1 - e\cos\eta){\rm d}\eta$, and ${\rm d}w' = (1 - e'\cos\eta'){\rm d}\eta'$. The following symmetries of  $\Psi(\scrr, \scrr')$ can be verified either by writing the integral in eqn.(\ref{rrint-disc}) in explicit form, or more simply by thinking about the gravitational energy between two Gaussian Rings:
\begin{itemize}
\item[{\bf P1:}] $\Psi$ is a real function which is even in both $L\,$ and $L'\,$.

\item[{\bf P2:}] The apsidal longitudes $g$ and $g'$ occur in $\Psi$ only in the combination $\vert g - g'\vert$. 

\item[{\bf P3:}] $\Psi$ is independent of both $g$ and $g'$ when one of the two Rings is circular (i.e. when $L = \pm I\,$ or $\,L' = \pm I'$ or both). 

\item[{\bf P4:}] $\Psi$ is invariant under the interchange of the 
two Rings. This can be achieved by any of the transformations:
$\{I,\,L\} \leftrightarrow\{I',\,L'\}\,$, or $\,g\leftrightarrow g'\,$ 
or both. In explicit form we have: $\,\Psi(I, L, g, I', L', g') = \Psi(I', L', g, I, L, g')$,  $\;\Psi(I, L, g, I', L', g') = \Psi(I, L, g', I', L', g)$, which implies that $\Psi(I, L, g, I', L', g') = \Psi(I', L', g', I, L, g)$. 
\end{itemize}

We now focus attention on disc eigenmodes and hence set $\Phi_1^{\rm tid} = 0$ in eqn.(\ref{ham1-axi}), so that the perturbed Hamiltonian $H_1 = \Phi_1$ is due only to the perturbed self--gravity.\footnote{However, in the unperturbed Hamiltonian $H_0$ of eqn.(\ref{hamaxi3}), the external axisymmetric contribution, $\Phi_0^{\rm ext}(I, L)$, need not be zero.} Since the coefficients in the LCBE of eqn.(\ref{lcbe-axi}) are independent of the variables, $g$ and $\tau$, it is natural to develop the perturbations in a Fourier series. We begin with the Fourier--expansion of the Ring--Ring interaction potential function $\Psi(\scrr, \scrr')$ in $g$ and $g'$. Using the fact that $\Psi$ depends only on the difference of the apsidal longitudes, we write:
\beq
\Psi(I, L, g, I', L', g') \;=\; \sum_{m=-\infty}^{\infty}\,C_m(I, L, I', L')
\exp{\left[{\rm i}m (g - g')\right]}\,.
\label{ring-ring-fou}
\eeq

\noindent
The symmetry properties of $\Psi$, given above in {\bf P1} to {\bf P4}, 
translate to the following properties of its Fourier coefficients:
\begin{itemize}
\item[{\bf F1:}] The $C_m$ are real functions that are even in $L$ and $L'$.

\item[{\bf F2:}] The $C_m$ are even in $m$: i.e. $\,C_m = C_{-m}\,$.

\item[{\bf F3:}] When either $L= \pm I\,$ or $\,L' = \pm I'\,$ or both, 
then $C_m = 0$ for all $m\neq 0$.

\item[{\bf F4:}] $C_m(I, L, I', L') = C_m(I', L', I, L)$.
\end{itemize}

\noindent
The perturbed DF for an eigenmode with azimuthal wavenumber $m$ and eigenfrequency $\omega$ can be written as: 
\beq
F_1(I, L, g, \tau)  \;=\; {\rm Re}\left\{F_{1m}(I, L)\exp{\left[{\rm i} \left(m g - \omega \tau\right)\right]}\right\}\,.
\label{pert1m}
\eeq
where $F_{1,-m} = F_{1m}^*\,$. To compute the corresponding perturbation in the self--gravity, we substitute eqns.(\ref{pert1m}) and (\ref{ring-ring-fou}) in  (\ref{pert-axi}). Then the perturbed Hamiltonian is 
\begin{eqnarray}
H_1(I, L, g, \tau)  &\;=\;& \Phi_1(I, L, g, \tau) \;=\; {\rm Re}\left\{\Phi_{1m}(I, L)\exp{\left[{\rm i} \left(m g - \omega \tau\right)\right]}\right\}\,,\nonumber\\[1em]
\Phi_{1m}(I, L) &\;=\;& 2\pi\int\,{\rm d}I'\,{\rm d}L'\,C_m(I, L, I', L')\,F_{1m}(I', L')\,.
\label{pertham1m}
\end{eqnarray}
 
\noindent
Substituting eqns.(\ref{pert1m}) and (\ref{pertham1m}) in the LCBE eqn.(\ref{lcbe-axi}), we obtain the following \emph{linear integral eigenvalue problem}:
\beq
\left[m\,\Omega(I, L)\;-\;\omega\,\right]\,F_{1m}\;=\; 
2\pi m\,\left(\frac{\p F_0}{\p L}\right)\,
\int\,{\rm d}I'\,{\rm d}L'\,\,C_m(I, L, I', L')\,F_{1m}(I', L')\,,
\label{eigen}
\eeq
\noindent 
which must be solved to determine the  eigenvalues $\omega$
and eigenfunctions $F_{1m}(I, L)$.

When $m=0$, eqn.(\ref{eigen}) implies that $\omega\,=\,0\,$. Therefore 
all DFs of the form $F_0(I, L)$ are secularly stable to axisymmetric 
perturbations. Moreover these perturbations are also time--independent, giving rise to nearby axisymmetric equilibria. It should be noted that 
this does not imply these discs are stable to fast axisymmetric instabilities: very cold discs violating the \citet{t64} criterion will be unstable to axisymmetric modes growing over times $\sim\tk$. As we discussed at the end of \S~3, MMS analysis is insensitive to these fast modes. Therefore when we consider Keplerian stellar discs, we understand that they must have velocity dispersions that are large enough to make them stable to fast modes.

For non--axisymmetric modes with $m\neq 0$, the integral eigenvalue problem is, in general, difficult to solve. However, we can prove a stability result for a class of stationary DFs. Let $F_0(I, L)$ 
be a strictly monotonic function of $L$ at fixed $I$. Then we can write
\beq
\frac{\p F_0}{\p L} \;=\; \sigma\left\vert\frac{\p F_0}{\p L}\right\vert
\;\neq\; 0\,,
\label{sigmadef}
\eeq

\noindent
where $\sigma = +1$ when $F_0$ is an increasing function of $L$, and 
$\sigma = -1$ when $F_0$ is a decreasing function of $L$.  Let us define a
new eigenfunction, $X_m(I, L)$, and a new kernel $A_m(I, L, I', L')$ by
\begin{subequations}
\begin{eqnarray}
X_m(I, L) &\;=\;& \left\vert\frac{\p F_0}{\p L}\right\vert^{-1/2}F_{1m}(I, L)\,,
\label{eigenfn-new}\\[1ex]
A_m(I, L, I', L') &\;=\;& 2\pi
\left\vert\frac{\p F_0}{\p L}\right\vert^{1/2}
C_m(I, L, I', L')
\left\vert\frac{\p F'_0}{\p L'}\right\vert^{1/2}\,,   
\label{kernel-new}
\end{eqnarray}
\end{subequations}
\medskip
where $F'_0 = F_0(I',L')$. Then, for $m\neq 0$,  the eigenvalue eqn.(\ref{eigen}) becomes
\beq
\left[\,\Omega(I, L) \;-\; \frac{\omega}{m}\,\right]X_m(I, L) \;=\;
\sigma\int\,{\rm d}I'\,{\rm d}L'\,\,A_m(I, L, I', L')\,X_m(I', L')\,.
\label{eigen-new}
\eeq 

\noindent
From {\bf F2} and {\bf F4} we see that the kernel, $A_m(I, L, I', L')$ is real, even in $m$, and symmetric under the interchange $\{I,\,L\}\leftrightarrow\{I',\,L'\}\,$. Hence $A_m$ is Hermitian, and all the eigenvalues $\omega$ are real for all $m\,$. Therefore we have proved: 

\medskip 
\noindent
$\bullet\,$ \emph{Stationary, axisymmetric discs with DFs $F_0(I, L)$ 
are neutrally stable to all secular perturbations when $\,\p F_0/\p L\,$ is of the same sign (either positive or negative) everywhere in its domain of support, $\,-I \leq L \leq I\,$ and $I_{\rm min} \leq I \leq 
I_{\rm max}\,$}.

\medskip
\noindent
These secularly stable DFs can have both prograde and retrograde populations of stars because $-I \leq L \leq I\,$. Hence the stability result applies 
to counter--rotating discs of stars with DFs that are strictly monotonic (either increasing or decreasing) functions of $L$, at fixed $I$. Depending on the sign of $\p F_0/\p L$, either prograde or retrograde stars dominate at every value of $I\,$. These secularly stable stellar discs have net rotation and include physically interesting cases, such as a secular analogue of the well--known Schwarzschild DF. We may also restate the above stability result as:

\medskip 
\noindent
$\bullet\;$ \emph{A necessary condition for $F_0(I, L)$ to be secularly unstable is that $\p F_0/\p L$ must vanish somewhere in its domain of support, $\,-I \leq L \leq I\,$ and $I_{\rm min} \leq I \leq I_{\rm max}\,$}.

\medskip
\noindent
\citet{jt12} derived the dispersion relationship for slow modes in a 
collisionless disc with a Schwarzschild DF --- using the epicyclic approximation and the WKB limit --- and showed that modes of all $m$ were stable. The stability result presented above is more general, and not restricted to a particular DF, or the epicyclic approximation or the WKB limit. Secular instabilities have been discussed in \citet{tre05, pps07}. \citet{tre05} considered non--rotating stellar discs described by DFs 
$F_0(I, L)$ that are even functions of $L\,$, and empty loss cones ($F_0/L$ nonsingular when $L\to 0\,$). Using a variational principle \citep{goo88} he argued that many of them are likely to be unstable to $m=1$ modes. 
\citet{pps07} considered mono--energetic counter--rotating discs dominated by nearly radial orbits, with DFs equivalent to $F_0 \propto \delta(I - I_0)A(L)\,$, and found that they are prone to loss cone instabilities of all 
$m\,$.

\section{Discussion}

The main results of this paper are:
\begin{itemize}
\item The full DF in 6--dim phase space, of a secularly evolving, collisionless Keplerian stellar system, is given by eqn.(\ref{fullsoln}). This the sum of an $O(1)$ secular DF in a reduced 5--dim phase space, and small fluctuations that remain of $O(\varepsilon)$ over secular times 
$\ts = \varepsilon^{-1}\tk\,$.

\item The slow time evolution of the secular (or Ring) DF $F(\scrr, \tau)$
in 5--dim $\scrr$--space is governed by the Ring CBE eqn.(\ref{cbe-des}), which includes the combined secular effects of the gravity of the MBH upto 1.5 PN order, the self--gravity of all the stars, and slowly varying external sources of arbitrary form.
\end{itemize}
The lower dimensional phase space results from the secular conservation of 
the semi--major axis of every stellar orbit. From this fully nonlinear, self--consistent formulation of secular dynamics follows a Secular Jeans theorem enabling the construction of Secular Equilibria, and problems concerning the Linear Response and Stability of these equilibria. Secular equilibria are easier to construct than full stellar dynamical equilibria. For instance, Jeans theorem implies that the exact DF of a stationary, spherical, non--rotating system is of the form $f(E, L)$. In order to determine $f$ as a function of the phase space coordinates, $\{\bfr, \bfu\}$, it is necessary to solve a self--consistent problem. However, the secular DF for the same system, $F(I, L)$, which differs from $f$ only by 
$O(\varepsilon)$, is known explicitly as a function of $\{\bfr, \bfu\}$. 

Linear secular stability of disc--like and spherical configurations has 
been studied earlier \citep{tre05, pps07}. Both begin with equilibria 
and linear perturbations of the full (non--secular) problem. Having 
formulated an eigenvalue equation for normal modes, they then take the 
slow mode --- i.e. $\omega \sim O(\varepsilon)$ --- limit. In contrast we 
have formulated a fully nonlinear secular theory, described by Gaussian
Ring dynamics. From this theory followed secular equilibria and development 
of secular linear perturbation theory. Broad conclusions were reached 
about evolutions of small perturbations in spherical non--rotating systems. 
\citet{tre05} has noted that flat systems are less simple than spherical 
systems due to a difficulty with defining inner--products, and the more 
complicated relation between mass distribution and potential. 
We considered perturbations of DFs describing axisymmetric, zero--thickness flat discs, and derived a secular integral eigenvalue problem. Even though the potential is related to the surface density in a more complicated manner, its symmetry properties were sufficient to let us prove a stability result for a class of physically interesting rotating DFs.

The fully nonlinear secular description of the Ring CBE is needed 
to go beyond linear theory. \citet{ts12} used a fully nonlinear secular CBE --- equivalent to the Ring CBE --- to discuss the counter--rotating instability, but they did not derive the secular equation 
from the full CBE, as we have done in this paper. \citet{ttk09} performed numerical simulations of secular dynamics, wherein each star is replaced by a Gaussian Ring that responds to the torques of all the other stars. They found that the counter--rotating instability in unstable discs saturates over times $\sim \ts$, and the discs settle into uniformly precessing equilibria. The in--plane instability unfolds into a triaxial one, which couples radial (or eccentricity) and vertical (or inclination) degrees--of--freedom, resulting in the dispersal of the lighter of the two counter--rotating components into a triaxial halo in which the other, more massive, precessing, lopsided disc is embedded. This process is the secular
analogue of ``violent relaxation'' \citep{l-b67} and forms a central problem in collisionless secular dynamics. The results obtained in this paper are
used directly in the papers to follow. In Paper~II we formulate a theory
of the resonant relaxation of general Keplerian stellar systems, in terms of the quasi--static evolution (due to gravitational collisions between Gaussian Rings) of the Ring DF through a sequence of collisionless equilibria, each of which must be dynamically stable. In Paper~III we 
apply this theory to study the physical kinetics of the resonant relaxation 
of axisymmetric discs.

\section*{Acknowledgments}
This work has benefited from a series of visits by S. Sridhar to Ras Beirut, made possible by the generous support of the Dean's Office and the hospitality of the Department of Physics, at the American University of Beirut. We thank Karamveer Kaur for a careful reading of an earlier draft.

\appendix
\section{Secular Dynamics of Stellar Orbits}

The full DF, $f(\scrd, t)$ of eqn.(\ref{fullsoln}), describes the 
collisionless behaviour of $N\gg 1$ stars around a MBH. The orbit
of a test star that belongs to this DF is governed by the Hamiltonian 
$H_{\rm org}$ of eqn.(\ref{ham0}). $H_{\rm org}$ is the sum of three 
terms: of these the one containing $\bfA_\bullet$ vanishes to $O(\varepsilon)$ accuracy (see eqn.\ref{azero}); the Kepler energy is $E_{\rm k}(I) = -1/2\left(GM_\bullet/I\right)^2$; and the self--gravitational potential is
\begin{eqnarray}
-\varepsilon\,GM_\bullet\int \frac{f(\scrd', t)}{\,\vert\bfr - \bfr'\vert\,}\,{\rm d}\scrd'
&\;=\;& 
-\varepsilon\,GM_\bullet\int F(\scrr', \varepsilon t )\,{\rm d}\scrr'
\oint \frac{{\rm d}w'}{2\pi}\frac{1}
{\,\vert\bfr - \bfr'\vert\,} \;+\; O(\varepsilon^2)
\nonumber\\[1em]
&\;=\;& \varepsilon\Phi(\scrr, \varepsilon t) \;+\; \varepsilon\sum_{n\neq 0}\varphi_n(\scrr, \varepsilon t)\exp{[{\rm i}n w]} \;+\; O(\varepsilon^2)\,,
\label{}
\end{eqnarray}

\noindent
where we have used eqns.(\ref{phiF}) and (\ref{phifou}). Hence, to
$O(\varepsilon)$ accuracy,
\beq
H_{\rm org}(\scrd, t) \;=\; E_{\rm k}(I) \;+\; \varepsilon\Phi(\scrr, \varepsilon t) \;+\; \varepsilon\sum_{n\neq 0}\varphi_n(\scrr, \varepsilon t)\exp{[{\rm i}n w]}\,,
\eeq

\noindent
depends on all six Delaunay variables, but has only slow variations with time. Therefore it is natural to want to use $\tau = \varepsilon t$ as the 
time variable, instead of $t$. The equations of motion are invariant under the rescaling, $\{t\,,\; H_{\rm org}\}\to\{\tau =\varepsilon t\,,\; H'_{\rm org} = \varepsilon^{-1}H_{\rm org}\}$. We have:
\beq
H'_{\rm org}(\scrd, \tau) \;=\; 
\frac{E_{\rm k}(I)}{\varepsilon} \;+\; \Phi(\scrr, \tau) \;+\; \sum_{n\neq 0}\varphi_n(\scrr, \tau)\exp{[{\rm i}n w]}\,.
\label{hamscal}
\eeq

\noindent
The next step makes use of a near--identity canonical transformation 
from old to new Delaunay variables, $\scrd\to\scrd'\equiv\{I', L', L'_z; 
w', g', h'\}$, through the mixed--variable generating function,
\beq
{\cal S}_0(I', L', L'_z; w, g, h; \tau)\;=\; I'w \,+\, L'g \,+\, L'_z h
\;+\; \varepsilon {\cal S}(I', L', L'_z; w, g, h; \tau)\,. 
\label{gf}
\eeq

\noindent
The old actions ($I, L, L_{z}$) and new angles ($w',g',h'$) are given by:
\begin{eqnarray}
I &\;=\;& I' \;+\; \varepsilon\frac{\p {\cal S}}{\p w}\,,\qquad\qquad
L \;=\; L' \;+\; \varepsilon\frac{\p {\cal S}}{\p g}\,,\qquad\qquad
L_{z} \;=\; L'_z \;+\; \varepsilon\frac{\p {\cal S}}{\p h}\,,
\nonumber\\[3ex]
w' &\;=\;& w \;+\; \varepsilon\frac{\p {\cal S}}{\p I'}\,,\qquad\qquad
g' \;=\; g \;+\; \varepsilon\frac{\p {\cal S}}{\p L'}\,,\qquad\qquad
h' \;=\; h \;+\; \varepsilon\frac{\p {\cal S}}{\p L'_z}\,.
\label{cantr}
\end{eqnarray}

\noindent
Hence $\scrd$ and $\scrd'$, the old and new variables, differ only by 
$O(\varepsilon)$. The new Hamiltonian governing the dynamics of $\scrd'$ 
is:
\begin{eqnarray}
H^{''}_{\rm org}(\scrd', \tau) &\;=\;& H'_0(\scrd, \tau) \;+\; \varepsilon\frac{\p {\cal S}}{\p \tau}
\nonumber\\[1em]
&\;=\;&  
\frac{E_{\rm k}(I)}{\varepsilon} \;+\; \Phi(\scrr, \tau) \;+\; \sum_{n\neq 0}\varphi_n(\scrr, \tau)\exp{[{\rm i}n w]} \;+\; \varepsilon\frac{\p {\cal S}}{\p \tau}\,.
\label{hamstar3}
\end{eqnarray}

\noindent
We need to express the right hand side in terms of the new variables 
$\scrd'$, and keep only the $O(1)$ terms. The first term can be expanded as:
\beq
\frac{E_{\rm k}(I)}{\varepsilon} \;=\;  \frac{1}{\varepsilon}E_{\rm k}\!\left(I' + \varepsilon\,\frac{\p {\cal S}}{\p w}\right) \;=\; \frac{E_{\rm k}(I')}{\varepsilon} \;+\; \Omega_{\rm k}(I')\frac{\p {\cal S'}}{\p w'} \;+\; O(\varepsilon)\,,\nonumber
\label{kepexp}
\eeq

\noindent
where ${\cal S}'(\scrd')= {\cal S}(I', L', L'_z; w', g', h'; \tau)$ is a function only of the new variables, obtained by replacing the old angles 
$(w, g, h)$ in ${\cal S}$ by the new ones $(w', g', h')$; this is legitimate, because, by eqns.(\ref{cantr}), the old and new variables 
differ only by $O(\varepsilon)$. For the same reason, 
in the two $O(1)$ terms, we can replace $\scrd$ by $\scrd'$. Then 
\beq
H^{''}_{\rm org}(\scrd', \tau) \;=\; 
\frac{E_{\rm k}(I')}{\varepsilon} \;+\; \Omega_{\rm k}(I')\frac{\p {\cal S}'}{\p w'} 
\;+\; \Phi(\scrr', \tau) \;+\; \sum_{n\neq 0}\varphi_n(\scrr', \tau)\exp{[{\rm i}n w']} \;+\; O(\varepsilon)\,.
\label{hamstar4}
\eeq
 
\noindent
We now choose ${\cal S}'$ such that all the $w'$--dependent terms on the 
right hand side cancel. This amounts to choosing the mixed variable
generating function,
\beq
{\cal S}(I', L', L'_z; w, g, h; \tau) \;=\;
-\frac{1}{\Omega_{\rm k}(I')}\,\sum_{n\neq 0}\frac{1}{{\rm i}n}\varphi_n(I', L', L'_z, g, h, \tau)\exp{[{\rm i}n w]}\,.
\label{gfsoln}
\eeq

\noindent
${\cal S}$ is well--defined, since $\Omega_{\rm k}(I')$ and $n$ are non--zero. Note that the degeneracy of the Kepler problem has prevented the occurrence of small denominators. Then the Hamiltonian for the dynamics 
of $\scrd'$ is:
\beq
H^{''}_{\rm org}(\scrd', \tau) \;=\; 
\frac{E_{\rm k}(I')}{\varepsilon} \;+\; \Phi(\scrr', \tau) \;+\; O(\varepsilon)\,.
\label{hamstar5}
\eeq

\noindent
with the corresponding equations of motion:
\begin{eqnarray}
\frac{{\rm d}I'}{{\rm d}\tau} &\;=\;&  O(\varepsilon)\,,\qquad \frac{{\rm d}w'}{{\rm d}\tau} 
\;=\; \frac{\Omega_{\rm k}(I')}{\varepsilon} \;+\; \frac{\p \Phi}{\p I'} \;+\; O(\varepsilon)\,;
\nonumber\\[1em]
\frac{{\rm d}L'}{{\rm d}\tau} &\;=\;& -\,\frac{\p \Phi}{\p g'} \;+\; O(\varepsilon)\,,\qquad  
\frac{{\rm d}g'}{{\rm d}\tau} \;=\; \frac{\p \Phi}{\p L'} \;+\; O(\varepsilon)\,;
\nonumber\\[1em]
\frac{{\rm d}L'_z}{{\rm d}\tau} &\;=\;& -\,\frac{\p \Phi}{\p h'} \;+\; O(\varepsilon)\,,\qquad  
\frac{{\rm d}h'}{{\rm d}\tau} \;=\; \frac{\p \Phi}{\p L'_z} \;+\; O(\varepsilon)\,.
\label{eom-new}
\end{eqnarray}

\noindent
The dominant behaviour of the orbital phase $w'$ is rapid circulation
at the Kepler orbital rate, $(\Omega_{\rm k}/\varepsilon)$. Superposed on this are two kinds of modulations; a slow $O(1)$ variation and a fast $O(\varepsilon)$ oscillation. The other five variables, $\scrr' = (I', L', L'_z, g', h')$, behave like Gaussian Rings with additional fast $O(\varepsilon)$ oscillations: to $O(1)$ accuracy, $I'$ is constant, while the other four variables, $(L', L'_z; g', h')$, vary over secular time scales with $\Phi(\scrr', \tau)$ acting as the Hamiltonian for their dynamics. This is just the dynamics of $\scrr'$ in the orbit--averaged self--gravitational potential. Since, by eqn.(\ref{cantr}), $\scrr$ and $\scrr'$ differ only by $O(\varepsilon)$, the old $\scrr$ variables also have $O(1)$ slow dynamics governed by the Hamiltonian $\Phi(\scrr, \tau)$ plus $O(\varepsilon)$
fast oscillations. Therefore a stellar orbit in a Keplerian stellar system 
can be thought of as a secularly evolving, ``noisy'' Gaussian Ring.$\hfill\square$
  
\end{document}